\documentclass{article}
\usepackage[OT1]{fontenc} 
\usepackage{arxiv}
\usepackage{graphicx} 

\usepackage{amsmath}
\usepackage{amssymb}
\usepackage{bm}
\usepackage{mathtools}
\usepackage{booktabs}
\usepackage{xcolor}
\usepackage{subcaption}
\usepackage{comment}
\usepackage{physics}
\usepackage{subcaption}
\usepackage{multirow}
\usepackage[ruled,vlined]{algorithm2e}
\newcommand{\argmin}{\operatorname{argmin}}
\usepackage{todonotes}

\usepackage{afterpage}
\usepackage{longtable}
\usepackage{array}
\usepackage{bbm}
\graphicspath{{./figures/}}
\usepackage{tikz}
\usepackage{booktabs}
\usepackage{natbib}
\usepackage{multirow}
\title{{\bf Deep Inverse Rosenblatt Transport for Structural Reliability Analysis }}

\author{
Aryan Tyagi \\
{\it
Department of Aerospace Engineering \& Engineering Mechanics}, \\
{\it
The University of Texas at Austin}, \\
Austin, TX, USA
\And
Jan N. Fuhg\thanks{{correspondence: \tt jan.fuhg@utexas.edu}} \\
{\it
Department of Aerospace Engineering \& Engineering Mechanics}, \\
{\it
\& The Oden Institute of Computational Science and Engineering,} \\
{\it
The University of Texas at Austin}, \\
Austin, TX, USA
}
\date{}
\newcommand{\Pf}{\mathbb{P}(\mathcal{F})}
\newcommand{\Pfest}{\hat{\mathbb{P}}(\mathcal{F})}
\newcommand{\Pfpost}{\mathbb{P}(\mathcal{F}\mid\mathbf{y})}
\newcommand{\Pfpostest}{\hat{\mathbb{P}}(\mathcal{F}\mid\mathbf{\tilde{y}})}

\begin{document}

\maketitle


\begin{abstract}

Accurately estimating the probability of failure in engineering systems under uncertainty is a fundamental challenge, particularly in high-dimensional settings and for rare events. Conventional reliability analysis methods often become computationally intractable or exhibit high estimator variance when applied to problems with hundreds of uncertain parameters or highly concentrated failure regions. In this work, we investigate the use of the recently proposed Deep Inverse Rosenblatt Transport (DIRT) framework for reliability analysis in solid mechanics. DIRT combines a TT decomposition with an inverse Rosenblatt transformation to construct a low-rank approximation of the posterior distribution, enabling efficient sampling and probability estimation in high-dimensional spaces. By representing the optimal importance density in the TT format, DIRT scales linearly in the input dimension while maintaining a compact, reusable surrogate of the target distribution. We demonstrate the effectiveness of the DIRT framework on three analytical reliability problems and one numerical example with dimensionality ranging from 2 to 250. Compared to established methods such as Bayesian updating with Subset Simulation (BUS-SuS), DIRT seems to lower the estimator variance while accurately capturing rare event probabilities for the benchmark problems of this study.

\end{abstract}

\section{Introduction}
In many engineering systems, uncertainties exist due to different factors, such as lack of data, incomplete knowledge, or assumed knowledge \citep{cascading,hybrid,seismic,bedford2001probabilistic}.  Reliability analysis takes these uncertainties into account and makes decisions based on the probability that the system fails \cite{beer2013reliability,faes2021engineering}. For rare event cases, the probability of failure can be on the order of $10^{-4}$ or lower. In structural reliability analysis, a limit state function is commonly used to define the boundary between safe and failed states of a system, typically in terms of a Quantity of Interest (QoI) that exceeds a critical threshold within a certain region of the input parameter space \cite{fiessler1979quadratic,gavin2008high,yuan2021efficient}. A classical example is that the maximum stress in a body should not exceed the yield strength of the material. Accurately estimating the probability of failure of a system has been an active area of research in the field of civil and structural engineering since the 1970s \citep{rackwitz1976note,ditlevsen1976collection,1970s}. 

Conventional methods in reliability analysis, such as FORM \citep{FORM_original} and SORM \citep{SORM_1} use use Taylor series expansions to approximate the limit state function near a design point, enabling efficient estimation of failure probabilities. However, historically, most methods in structural reliability have focused on computing the probability of failure under the assumption of complete knowledge of the input parameter distributions, that is, evaluating the probability of failure based on the prior probabilistic model \citep{prior1,prior2,prior3,prior4,prior5}. Estimates of failure probabilities can be improved by using observational data of the system response, leading to a Bayesian inference task \cite{martz2014b}. 

In the Bayesian approach, the posterior distributions of the model parameters are initially estimated using noisy observation data \citep{bayesian_1,bayesian_2}. These posterior distributions are then used, typically via sampling, to compute the updated (posterior) probability of failure by integrating over the remaining parameter uncertainty. This can be inaccurate, since the approximation of the posterior distribution itself may be erroneous. Also, in the context of reliability analysis, even after the posterior has been approximated, the sample size still needs to be large enough to draw failure samples.

To address these limitations, an alternative approach has been proposed that draws an analogy between Bayesian updating and structural reliability analysis \citep{BUS}. This formulation, known as Bayesian Updating with Structural Reliability (BUS) Method, enables the use of established reliability techniques to perform Bayesian inference more efficiently. By leveraging tools from structural reliability, BUS avoids direct posterior sampling and can yield more accurate and computationally efficient estimates, particularly in problems involving rare events. 
In this context, Importance Sampling (IS) can be employed \citep{IS1,IS2,IS3}. IS draws samples from a proposal distribution instead of the actual (often difficult to sample from) distribution of the random variables, making it possible to draw more samples from the failure region. The samples from the proposal distribution are then assigned \textit{importance weights} to compute an estimate for the failure probability. If designed properly, IS can have lower variance compared to Markov Chain Monte Carlo (MCMC).

The BUS method can be used in conjunction with other efficient sampling methods, such as Subset Simulation (SuS) \citep{Subset_original}, to draw more frequent samples from the failure region. However, the developed BUS-SuS method \citep{BUS} can still be computationally expensive for rare events since the number of calls required to compute the failure probability increases as the the failure probability decreases.


Nevertheless, Uribe et al. \cite{BUS_KLE} employed BUS-SuS for random fields represented by a Karhunen-Loeve (KL) expansion to find the posterior probability of failure in high dimensions. Their results show Coefficient of Variation (CoV) values in the $0.3 - 0.8$ range for the updated probabilities for $d>20$, which is much higher than the desired CoV ($\sim$0.05) for applications such as structural reliability \citep{CoV5}. 

To overcome the limitations of classical reliability methods in high-dimensional settings, we investigate a recently developed framework that employs TT decomposition \citep{TT_original} to store the IS proposal distribution. This framework, termed Deep Inverse Rosenblatt Transport (DIRT),  was originally proposed in the context of Bayesian inference for high-dimensional distributions \citep{DIRT2021} in a computational math context. It allows for efficient sampling and probability estimation in situations where both the target posterior and the failure region are concentrated in a small subset of the input space. 
The key idea is to represent the optimal importance density as a high-dimensional tensor and approximate it using TT decomposition. The TT format enables us to compress and evaluate complex multivariate functions efficiently, with memory and computational cost that scale linearly with dimension \citep{approxsampling_TT}. This property of the TT decomposition has proven to be useful when applied to high-dimensional problems \citep{TT1,TT2,TT3}.  
Once the posterior is represented in the TT format, we use it to construct a TT-based inverse Rosenblatt transformation, which allows us to generate (approximately) independent samples from the posterior. These samples are then used in a double-proposal IS estimator to compute the posterior probability of failure. This IS scheme avoids direct evaluation of the normalization constant of the posterior and is well-suited to problems where evaluating the forward model is computationally expensive.

While the DIRT framework was originally developed for Bayesian PDE inverse problems, here we apply, test, and compare it to state-of-the-art methods in structural reliability analysis. This involves (i) embedding limit state functions and smoothing techniques into the TT construction, (ii) benchmarking against reliability-specific algorithms (BUS-SuS, CE), and (iii) demonstrating scalability on structural mechanics case studies up to 250 dimensions.
 We demonstrate that DIRT is capable of accurately estimating rare event probabilities even in high-dimensional settings, where other existing methods like BUS-SuS, IS with Gaussian mixtures, or FORM/SORM might either become intractable or unreliable. We investigate the feasibility of the framework on four numerical examples ranging from two dimensions up to multiple hundred dimensions. Our results indicate that DIRT reduces the number of forward model evaluations significantly while maintaining a reasonable Coefficient of Variation (CoV) of the updated failure probabilities, making it suitable for large-scale structural systems with complex physics and uncertainty.

The remainder of this work is structured as follows. In Section 2, we provide background on probability of failure, TT decomposition, and the DIRT framework. In Section 3, we test the framework on numerical experiments where we compute the posterior Pf for four structural reliability problems. The main
The findings of the study are summarized in Section 4.

\section{Background}
Let $\mathbf{X}\in\mathbb{R}^{d}$ denote the vector of uncertain input parameters characterized by the probability density function (PDF) $\pi(\mathbf{x})$, where $\mathbf{x}$ is a realization of $\mathbf{X}$. In structural reliability, uncertain parameters can include external loads, material properties, or geometric tolerances. The system's safety is described by the limit state function (LSF) denoted as $g(\mathbf{x})$. The LSF distinguishes the system's safe states ($g(\mathbf{x}) > 0$) from its failure states ($g(\mathbf{x}) \leq 0$). The failure domain can then be defined as the subset $\Omega_\mathcal{F}$ of $\Omega$, i.e.,
\begin{equation}
    \Omega_\mathcal{F} = \{\mathbf{x} \in \Omega : g(\mathbf{x}) \leq 0 \} \, .
\end{equation}

The probability of failure is then computed as
\begin{equation}\label{pf_expectation}
    \mathbb{P}
    ({\mathcal{F}}) = \mathbb{P}[g(\mathbf{x})\leq0] = \int_{\mathbb{R}^d} \mathbb{I}_{\Omega_\mathcal{F}}\pi(\mathbf{x}) \, d\mathbf{x}
    \quad \text{or equivalently} \quad         
    \Pf = \mathbb{E}_{\pi}\{\mathbb{I}_{\Omega_\mathcal{F}}(\mathbf{x})\},
\end{equation}
where $\mathbb{I}_{\Omega_\mathcal{F}}$ is the indicator function for the failure domain, equal to $1$ if the system fails and $0$ otherwise.  
To compute the expectation in Eq.~\eqref{pf_expectation} numerically, one can use crude Monte Carlo sampling to get an estimate for $\mathbb{P}(\mathcal{F})$:
\begin{equation}
\label{MCS_Pf}
    \hat{\mathbb{P}}(\mathcal{F}) \approx \frac{1}{N} \sum_{i=1}^{N} \mathbb{I}_{\Omega_\mathcal{F}}(\mathbf{x}^i), 
    \quad \mathbf{x}^i \overset{\text{i.i.d.}}{\sim} \pi(\mathbf{x}) \, ,
\end{equation}
over a large number of samples $N$.  
However, this is difficult for structural problems for several reasons, namely:
\begin{itemize}
    \item The LSF $g(\mathbf{x})$ can be computationally expensive to evaluate since it usually requires a Finite Element Method (FEM) simulation for each forward evaluation.
    \item Sampling directly from the density function $\pi(\mathbf{x})$ in Eq.~\eqref{MCS_Pf} can be difficult.
    \item The number of samples required to obtain a good estimate of $\Pf$ increases as $\Pf$ decreases. This becomes particularly challenging for rare events where $\Pf \approx 10^{-4}$. 
\end{itemize}

Importance sampling is frequently used in structural reliability to improve sampling efficiency by using a proposal PDF $p$ that over-represents the failure domain $\Omega_\mathcal{F}$. Also, it is assumed that it is easier to sample from $p$ as compared to $\pi$. IS therefore reduces the number of FEM simulations required to compute $\Pfest$. In particular, it allows computing $\mathbb{E}_{\pi}\{\mathbb{I}_{\{ g(\mathbf{x}\leq0\}}\}$ without drawing samples from $\pi$ directly. The IS estimator for the failure probability can be written as

\begin{equation}
    \Pfest = \frac{1}{N} \sum_{i=1}^{N} \mathbb{I}_{\{g(\mathbf{x^i})\leq0\}}\frac{ \pi(\mathbf{x}^i)}{p(\mathbf{x}^i)}, 
    \quad \mathbf{x}^i \overset{\text{i.i.d.}}{\sim} p \, .
\end{equation}

In structural reliability, it is often useful to update the failure probability when new information becomes available through sensor measurements, inspection data, or monitoring of system responses. This can be formulated using a Bayesian framework where the prior distribution $\pi_0(\mathbf{x)}$ describes the uncertainty in input parameters through previous knowledge, and the observed data $\mathbf{y}$ is incorporated in a likelihood function $\mathcal{L}^\mathbf{y}(\mathbf{x})$. The posterior distribution of the uncertain parameters can then be represented as
 
\begin{equation}
\pi^{\mathbf{y}}(\mathbf{x}) 
= \frac{1}{Z} \, \mathcal{L}^{\mathbf{y}}(\mathbf{x}) \, \pi_0(\mathbf{x}), 
\quad Z = \mathbb{E}_{\pi_0} \{\mathcal{L}^{\mathbf{y}}(\mathbf{x})\},
\end{equation}
where $Z$ is the normalization constant.

Under the posterior distribution, the failure probability conditioned on the observed data then reads
\begin{equation}
\Pfpost = 
\mathbb{E}_{\pi^{\mathbf{y}}}\{\mathbb{I}_{\{g(\mathbf{x})\leq0\}}\} 
= \frac{1}{Z} \int_{\Omega} \mathbb{I}_{\{g(\mathbf{x})\leq0\}} \, \mathcal{L}^{\mathbf{y}}(\mathbf{x}) \, \pi_0(\mathbf{x}) \, d\mathbf{x} \, .
\end{equation}

Following the IS formulation in Ref. \cite{DIRTRareEvent2022}, $\Pfpostest$ can be expressed as a ratio of two IS estimators. 

By drawing i.i.d. samples from proposal PDFs $\pi_{1}$ and $\pi_{2}$, we obtain the following importance sampling estimators:

\begin{equation}
\hat{Q}_{\pi_1,N} = \frac{1}{N} \sum_{i=1}^N \mathbb{I}_{\{g(\mathbf{x^i})\leq0\}}\frac{\mathcal{L}^{\mathbf{y}}(\mathbf{x}_{\pi_1}^i) \, \pi_0(\mathbf{x}_{\pi_1}^i)}{\pi_1(\mathbf{x}_{\pi_1}^i)}, 
\quad 
\hat{Z}_{\pi_2,N} = \frac{1}{N} \sum_{i=1}^N \frac{\mathcal{L}^{\mathbf{y}}(\mathbf{x}_{\pi_2}^i) \, \pi_0(\mathbf{x}_{\pi_2}^i)}{\pi_2(\mathbf{x}_{\pi_2}^i)} \, .
\end{equation}

This leads us to the estimator for the posterior failure probability $\Pfpostest$:
\begin{equation}
\label{eq:ratio_estimator}
\Pfpostest := \frac{\hat{Q}_{\pi_1,N}}{\hat{Z}_{\pi_2,N}} \, .
\end{equation}

The rest of this section explains the tools and methods used in the framework, namely, the TT Decomposition and the Inverse Rosenblatt Transformation. The complete framework is summarized in Figure~\ref{fig:flowchart}. For more details, the reader is referred to Refs. \citep{DIRT2021,DIRTRareEvent2022}.  

\subsection{Tensor Train Decomposition}
The Tensor Train (TT) decomposition provides a scalable, low-rank representation for multivariate functions, enabling efficient evaluation and sampling in high-dimensional spaces.  
To represent a $d$-dimensional PDF $\pi$ in TT format, we discretize $\pi$ on a $n_1 \times n_2 \times \ldots \times n_d$ grid, resulting in a tensor  

\begin{equation}
    \hat{\pi} \in \mathbb{R}^{n_1 \times n_2 \times \ldots \times n_d}.
\end{equation}

The TT factorization of $\hat{\pi}$ consists of cores $\{\pi^{(k)}\}_{k=1}^d$, where  
\begin{equation}
\pi^{(k)} \in \mathbb{R}^{r_{k-1} \times n_k \times r_k}, 
\quad r_0 = r_d = 1,
\end{equation}

such that, for any multi-index $(i_1, \ldots, i_d)$,  
\begin{equation}
    \hat{\pi}(i_1, \ldots, i_d) = \pi^{(1)}(i_1) \, \pi^{(2)}(i_2)  \ldots  \pi^{(d)}(i_d),  
\end{equation}

where $\pi^{(k)}(i_k) \in \mathbb{R}^{r_{k-1} \times r_k}$ denotes the $i_k$-th slice of the $k$-th TT core, and $(r_0, \ldots, r_d)$ are the TT ranks.  
If $r_k \le r$ and $n_k \le n$ for all $k$, the storage cost of $\hat{\pi}$ in TT format is $\mathcal{O}(dn r^2)$, which scales linearly in $d$.  
By contrast, storing the full tensor requires $\mathcal{O}(n^d)$ memory. This compression advantage diminishes if TT ranks become large.

\subsection{TT-Cross Algorithm}
A TT representation can be constructed via truncated singular value decomposition, but this requires knowledge of all tensor entries, negating the benefits of low-rank approximations. Cross approximation methods address this by using only selected rows and columns, drastically reducing the number of evaluations of the target tensor. In the DIRT framework, the Tensor Train Cross-Approximation (TT-Cross) algorithm \cite{TT_Cross} is used to approximate the target multivariate PDF. 

\subsubsection{Skeletal Decomposition for Matrices}
The TT-cross algorithm builds upon the concept of skeleton decomposition for matrices, so it is appropriate to provide the reader with some intuition about the skeleton decomposition here.   
Let $\mathbf{A} \in \mathbb{R}^{m \times n}$ be a rank-$r$ matrix. Its skeleton decomposition is:
\begin{equation}
    \mathbf{A} \approx \mathbf{C} \, \mathbf{U}^{+} \, \mathbf{R},
\end{equation}
where $\mathbf{C} \in \mathbb{R}^{m \times r}$ contains $r$ selected columns of $\mathbf{A}$,  
$\mathbf{R} \in \mathbb{R}^{r \times n}$ contains $r$ selected rows, and  
$\mathbf{U} \in \mathbb{R}^{r \times r}$ is the intersection submatrix between $\mathbf{C}$ and $\mathbf{R}$.  
The superscript $(\cdot)^{+}$ denotes the pseudoinverse.  
If $r \ll \min(m,n)$, storing $\mathbf{C}$, $\mathbf{R}$, and $\mathbf{U}$ requires only $(m + n - r)r$ elements, much less than $mn$.  
The efficiency of this method depends critically on the choice of rows and columns, often determined via the maximum-volume (\texttt{maxvol}) principle \cite{maxvol_ref}. The \texttt{maxvol} algorithm selects a subset of rows or columns whose intersection submatrix has the largest possible volume (absolute value of determinant). This ensures the submatrix $\mathbf{U}$ is well-conditioned, which improves numerical stability and accuracy in reconstructing the remaining entries from the selected ones.

\begin{figure}
    \centering
    \includegraphics[width=0.85\linewidth]{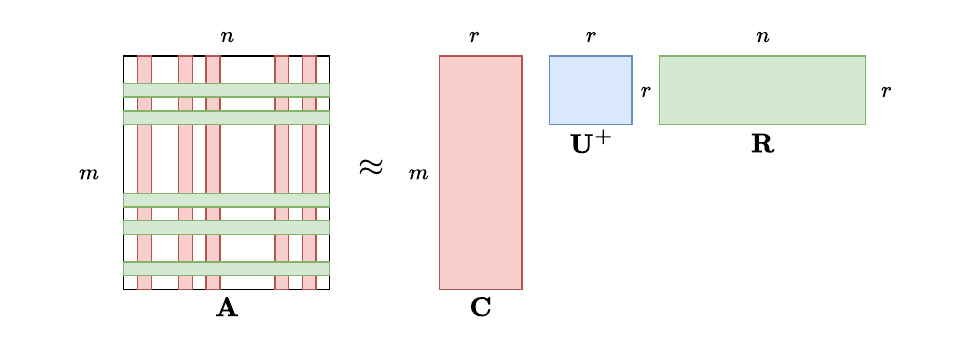}
    \caption{Skeleton decomposition of a matrix $\mathbf{A}$ into $\mathbf{C}$, $\mathbf{U}$, and $\mathbf{R}$.}
    \label{fig:CUR}
\end{figure}
\subsubsection{Cross Approximation for Tensor Train Decomposition}
The TT-cross algorithm generalizes skeleton decomposition to tensors.  
For $\hat{\pi} \in \mathbb{R}^{n_1 \times n_2 \times \ldots \times n_d}$, define the $k$-th unfolding:
\begin{equation}
    \hat{\pi}^{(k)} \in \mathbb{R}^{(n_1 \cdots n_k) \times (n_{k+1} \cdots n_d)},    
\end{equation}

where the first $k$ indices form the rows and the remaining indices form the columns. The algorithm relies on two sets of index selections $I_k$ and $J_k$. $I_k$ (left index) denotes the positions of rows and $J_k$ (right index) denotes the columns in the $k$-th unfolding. 

The TT-cross algorithm proceeds recursively:
\begin{itemize}
    \item \textbf{Forward Sweep:} Starting from $k = 1$, the tensor $\hat\pi$ is unfolded along the $k$-th dimension and a skeleton decomposition is performed
    \begin{equation}
        \hat{\pi}^{(k)} \approx C_k \, U_k^{+} \, R_k,    
    \end{equation}
    
    where $C_1 \in \mathbb{R}^{n_1 \times r_1}$, $R_1 \in \mathbb{R}^{r_1 \times (n_2 \cdots n_d)}$, and $U_1 \in \mathbb{R}^{r_1 \times r_1}$ is the intersection matrix.  
    The rows and columns are selected using the \texttt{maxvol} criterion to ensure $U_1$ is well-conditioned. In the forward sweep, the left index $I_k$ is updated.
    \item \textbf{Backward Sweep:} Starting from $k = d$, the unfolding is repeated in the reverse direction and the right index $J_k$ is updated using the \texttt{maxvol} algorithm. 
\end{itemize}

The alternating forward and backward sweeps produce TT-cores $\{\hat\pi^{(k)}\}_{k=1}^d$. The process is repeated until the maximum iterations are reached or the relative change between successive approximations falls below the prescribed threshold $\delta$. 

In our implementation, to obtain the continuous TT approximation of $\hat\pi$, we use a linear interpolation of TT blocks given as

\begin{equation}
\pi^{(k)}(x_k) =
\frac{x_k - x_k^{i_k}}{x_k^{i_k+1} - x_k^{i_k}} \, \hat{\pi}^{(k)}(i_k+1)
+ \frac{x_k^{i_k+1} - x_k}{x_k^{i_k+1} - x_k^{i_k}} \, \hat{\pi}^{(k)}(i_k).    
\end{equation}

The accuracy and efficiency of the TT-cross algorithm described above depend on:
\begin{enumerate}
    \item The univariate grids $\{x_k^{(i_k)}\}_{i_k=1}^{n_k}$ in each dimension.
    \item The choice of the initial right index sets $J_k$.
\end{enumerate}

Following Ref. \cite{approxsampling_TT}, we use uniform tensor-product grids and generate initial right index sets from independent realizations of the tractable reference distribution.

\begin{algorithm}[H]
\caption{TT-cross algorithm for TT approximation of $\pi$.}
\KwIn{Initial right index sets $J_k$, tolerance $\delta > 0$, maximum iterations $\mathrm{iter}_{\max}$.}
\KwOut{TT cores $\{\pi^{(k)}\}_{k=1}^d$ approximating $\pi(\mathbf{x})$.}
\While{$\mathrm{iter} < \mathrm{iter}_{\max}$ \textbf{and} $\|\tilde{\pi}^{(\mathrm{iter})} - \tilde{\pi}^{(\mathrm{iter}-1)}\| > \delta \|\tilde{\pi}^{(\mathrm{iter})}\|$}{
    \For{$k = 1, 2, \ldots, d$ \tcp*[f]{Forward sweep}}{
        Compute unfolding $\hat{\pi}(I_k, i_k; J_k)$ of size $r_{k-1} n_k \times r_k$\;
        Update $I_{k+1}$ via \texttt{maxvol} and truncate\;
    }
    \For{$k = d, d-1, \ldots, 1$ \tcp*[f]{Backward sweep}}{
        Compute unfolding $\hat{\pi}(I_k; i_k, J_k)$ of size $r_{k-1} \times n_k r_k$\;
        Update $J_{k-1}$ via \texttt{maxvol} and truncate\;
    }
}
\end{algorithm}

\subsection{Inverse Rosenblatt Transformation}
The random variables in structural reliability problems can often have complicated correlations and non-Gaussian PDFs. This makes it difficult to sample from the joint PDFs directly. To overcome this, the Inverse Rosenblatt Transformation (IRT) constructs a map $\mathcal{M}$ that transforms a set of independent uniform random variables into samples from a given multivariate probability distribution. It relies on the fact that any joint distribution can be decomposed into a sequence of conditional distributions. 

Given a joint PDF $\pi(x_1, \ldots, x_d)$, the Rosenblatt transformation maps a point $\mathbf{x} = (x_1, \ldots, x_d)$ to a point $\mathbf{u} = (u_1, \ldots, u_d) \in [0,1]^d$ via the sequence of conditional cumulative distribution functions (CDFs):
\begin{equation}
    \begin{aligned}
        u_1 &= \mathcal{M}_1(x_1) = \int_{-\infty}^{x_1} \pi_1(y_1) \, dy_1, \\
u_2 &= \mathcal{M}_2(x_2 \mid x_1) = \int_{-\infty}^{x_2} \pi_2(y_2 \mid x_1) \, dy_2, \\
&\ \vdots \\
u_d &= \mathcal{M}_d(x_d \mid x_1, \ldots, x_{d-1}) = \int_{-\infty}^{x_d} \pi_d(y_d \mid x_1, \ldots, x_{d-1}) \, dy_d,
    \end{aligned}
\end{equation}
where $\pi_1(y_1)$ is the marginal PDF of the first variable and $\pi_k(y_k\mid x_1,...,x_{k-1})$ is the conditional PDF of the $k$-th random variable given the preceding random variables. 
The inverse map takes $\mathbf{u} = (u_1, \ldots, u_d)$ and sequentially solves for $\mathbf{x} = (x_1, \ldots, x_d)$ by inverting the above CDF relations:
\begin{equation}
    \begin{aligned}
        x_1 &= \mathcal{M}_1^{-1}(u_1), \\
x_2 &= \mathcal{M}_2^{-1}(u_2 \mid x_1), \\
&\ \vdots  \\
x_d &= \mathcal{M}_d^{-1}(u_d \mid x_1, \ldots, x_{d-1}).
    \end{aligned}
\end{equation}

Thus, independent uniform or Gaussian samples can be transformed into realizations of the input random variables that follow the target joint distribution. When used within structural reliability, these transformed samples can be directly evaluated in the LSF $g(\mathbf{x})$ to compute failure probabilities. 

\paragraph{Need for TT representation.}
Even when $\hat\pi$ has been approximated, direct evaluation of the conditional marginals and CDFs scales poorly with $d$. Representing $\hat\pi$ in TT format allows these operations to be carried out efficiently, with cost scaling linearly in $d$ for bounded TT ranks. This makes IRT feasible for high-dimensional distributions. The details of computing the marginal and conditional CDFs given an approximation of the target PDF in TT format are given in Appendix \ref{app1}.

\subsection{Multi-Layer Construction of the Inverse Rosenblatt Transformation}
In practice, the target density $\pi(\mathbf{x})$ for rare-event probability estimation is often highly concentrated in a small region of the parameter space. This makes a direct, single-step IRT construction difficult, 
since approximating $\pi(\mathbf{x})$ in one shot can require very large TT ranks and fine discretization, leading to a significant computational burden. 
 
To address this issue, the DIRT framework adopts a multi-layer construction strategy. Instead of building the full transport in one step, the mapping from the reference density $\rho$ to the target density $\pi$ is achieved gradually through a sequence of $L$ intermediate transformations:

\begin{equation}
\mathcal{T}^{(l)} = \mathcal{M}^{(1)} \circ \mathcal{M}^{(2)} \circ \cdots \circ \mathcal{M}^{(l)},
\quad l=1,\ldots,L,    
\end{equation}

where each layer incrementally adapts the distribution toward the target.
At layer $l$, we define an intermediate unnormalized density $\phi^{(l)}$ using a tempering approach,
\begin{equation}
    \phi^{(l)}(\mathbf{x}) \propto \pi(\mathbf{x})^{\beta_l},    
\end{equation}

where $\beta_l \in (0,1]$ controls the gradual transition from the reference distribution ($\beta_0=0$) to the full target distribution ($\beta_L=1$). This stabilizes the construction by smoothing the progression across layers.

The normalized intermediate density is given by
\begin{equation}
    \varphi^{(l)}(\mathbf{x}) = \frac{\phi^{(l)}(\mathbf{x})}{\omega^{(l)}}, 
    \quad \omega^{(l)} = \int_{\mathcal{X}} \phi^{(l)}(\mathbf{x})\,dx
\end{equation}

The composite map $\mathcal{T}^{(l)}$ is then constructed so that the pushforward of $\rho$ under 
$\mathcal{T}^{(l)}$ approximates $\varphi^{(l)}$,
\begin{equation}
    \{\mathcal{T}^{(l)}\}_{\sharp}\rho \approx \varphi^{(l)}.    
\end{equation}

To proceed to the next layer, we extend the map by including a new transformation 
$\mathcal{M}^{(l+1)}$, so that
\begin{equation}       
    \mathcal{T}^{(l+1)} = \mathcal{T}^{(l)} \circ \mathcal{M}^{(l+1)},
\end{equation}
and the pushforward of $\rho$ under $\mathcal{T}^{(l+1)}$ approximates $\varphi^{(l+1)}$. 

Equivalently, in terms of pullbacks, this condition can be written as
\begin{equation}
    \{\mathcal{M}^{(l+1)}\}_{\sharp} \varphi^{(l)} \approx \varphi^{(l+1)}.    
\end{equation}

This layered construction is advantageous because it avoids the difficulty of approximating the full target density directly. Instead, each step only needs to approximate a smoother intermediate density, which is computationally more tractable and easier to represent in TT format.

\begin{figure}[t]
    \centering
    \includegraphics[width=1.1\linewidth]{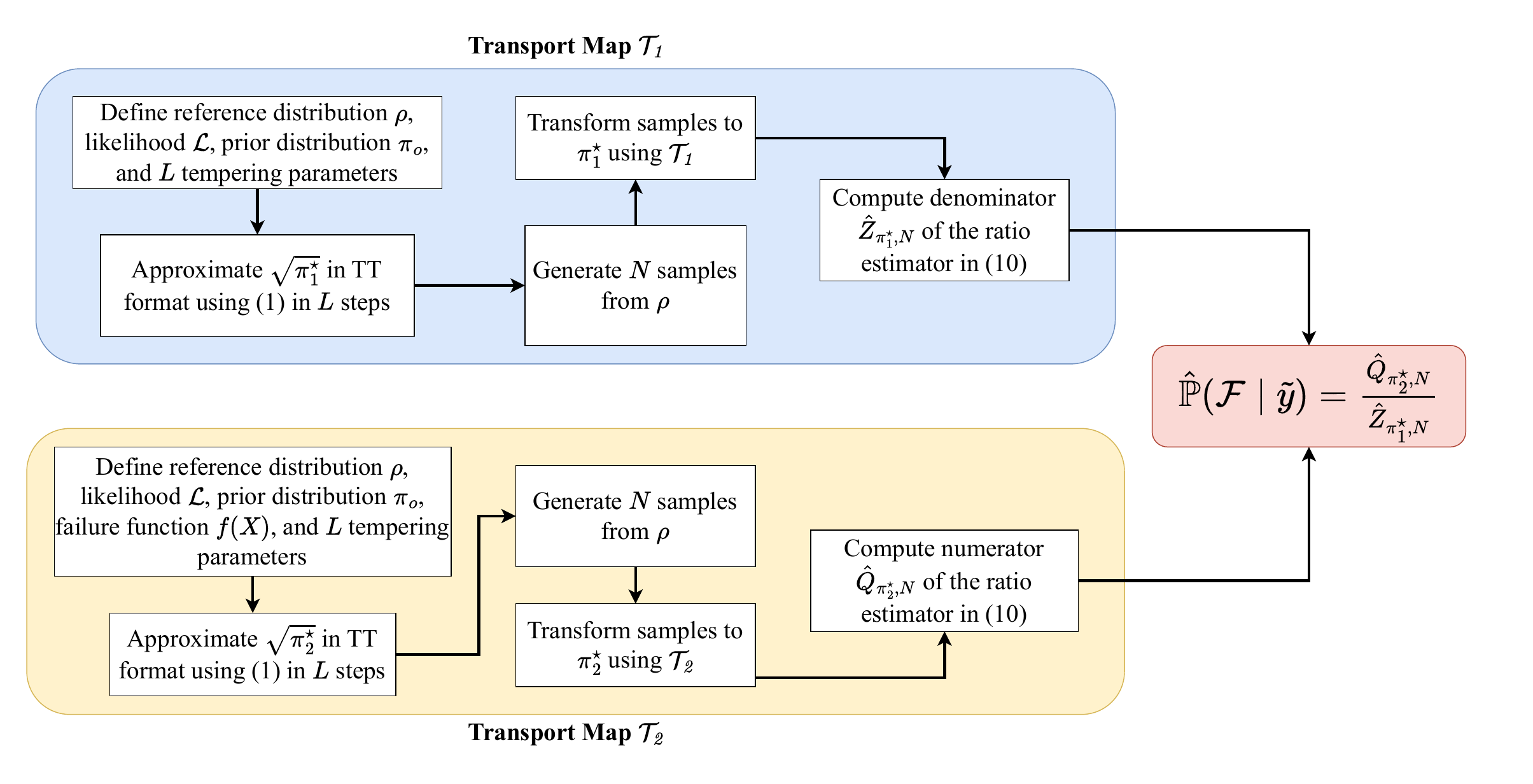}
    \caption{Summary of the Overall Framework for Estimating Probability of Failure. Both Transport Maps are constructed in parallel}
    \label{fig:flowchart}
\end{figure}

The indicator of the failure domain $\mathbb{I}_{\{g(\mathbf{x})\leq0\}}$ is discontinuous at the limit state surface $g(\mathbf{x})=0$. This leads to difficulties in the TT approximation of the density function since it leads to high TT ranks and makes it difficult to discretize $\pi^{\star}$ efficiently. To mitigate this problem, the sharp indicator function is replaced by a smooth surrogate controlled by a user-defined parameter $\gamma$, which controls the trade-off between bias and variance in the estimation of $\Pfpostest$ and must be tuned accordingly. Specifically, DIRT approximates the indicator function by a sigmoid function that converges to $\mathbb{I}_{\{g(\mathbf{x})\leq0\}}$ as $\gamma\rightarrow\infty$

\begin{equation}
\label{failure_cont}
    \mathbb{I}_{\{g(\mathbf{x})\leq0\}}^{\gamma} = s_\gamma(g(\mathbf{x}))
\end{equation}

where $s_\gamma$ is the smoothing function given as

\begin{equation}
\label{sigmoid}
    s_\gamma(g(\mathbf{x})) = [1 + \exp\{\gamma(g(\mathbf{x)}\}]^{-1} \, .
\end{equation}

Finally, to estimate the updated or posterior probability of failure, we construct two composite maps $\mathcal{T}_1$ and $\mathcal{T}_2$ to approximate the optimal importance sampling distributions $\pi_1^\star$ and $\pi_2^\star$, respectively. Then, we draw samples from the reference distribution $\rho$ and transform these samples to the two target distributions. This leads to the following ratio estimator after accounting for the transformation of the samples drawn from $\rho$.

\begin{equation}
\Pfpostest = \frac{\hat{Q}_{\pi_1^\star, N}}{\hat{Z}_{\pi_2^\star, N}}, \quad 
\hat{Q}_{\pi_1^\star, N} = \frac{1}{N} \sum_{i=1}^N w_Q(\mathbf{u}_{\pi_1}^i), \quad 
\hat{Z}_{\pi_2^\star, N} = \frac{1}{N} \sum_{i=1}^N w_Z(\mathbf{u}_{\pi_2}^i), \quad 
\mathbf{u}_{\pi_1}^i, \mathbf{u}_{\pi_2}^i \sim \rho,
\end{equation}

\noindent where

\begin{equation}
w_Q(U) = \mathbb{I}_{\{g(\mathbf{\mathcal{T}_{1}(\mathbf{u})})\leq0\}}^{\gamma}\frac{\mathcal{L}^y\left\{\mathcal{T}_{1}(\mathbf{\mathbf{u}})\right\} \pi_0\left\{\mathcal{T}_{1}(\mathbf{u})\right\}}{\pi_1^\star\left\{\mathcal{T}_{1}(\mathbf{u})\right\}}, \quad 
w_Z(U) = \frac{\mathcal{L}^y\left\{\mathcal{T}_{2}(\mathbf{u})\right\} \pi_0\left\{\mathcal{T}_{2}(\mathbf{u})\right\}}{\pi_2^\star\left\{\mathcal{T}_{2}(\mathbf{u})\right\}}.
\end{equation}

It is important to note here that the DIRT framework approximates the square root of the optimal importance density, $\sqrt{\pi^\star}$, in the TT format rather than directly decomposing $\pi^\star$. This choice is motivated by the following considerations in Ref. \cite{DIRT2021}:
\begin{itemize}
    \item Although $\pi^\star$ is non-negative, its TT approximation can exhibit negative values, which in turn may cause the IRT map to lose \textit{monotonicity}. Since the IRT is constructed via conditional CDFs, monotonicity is essential to ensure that the map is a valid and invertible transport between probability measures.
    \item Taking the modulus of $\pi^\star$ avoids negativity but introduces discontinuities, thereby reducing the smoothness and accuracy of the resulting IRT. 
\end{itemize}

To overcome these issues, DIRT instead represents $\sqrt{\pi^\star}$ in the TT format, which ensures both non-negativity and smoothness:
\begin{equation}
\sqrt{\pi^{\star}(x)} \approx  \pi^{\star(1)}(x_1) \cdots \pi^{\star(k)}(x_k) \cdots \pi^{\star(d)}(x_d),
\end{equation}
where each core $\pi^{\star(k)}(x_k)$ has dimensions $r_{k-1}\times r_k$.

\paragraph{Final hyperparameters that need to be chosen in DIRT}

In summary, the user has to choose the tempering parameters ($\beta^l$), number of layers ($L$), maximum rank of the TT approximation ($r$), and the smoothing parameter in the failure function ($\gamma$).

\section{Numerical Examples}
The following sections contain numerical examples starting from low-dimensional reliability problems and moving up to higher dimensions to test the proposed method for a wide range of problems in structural reliability. The first example is used to show the performance of DIRT in computing the prior probability of failure. The rest of the examples deal with posterior probabilities of failure by first inferring the posterior distribution of the unknown parameters. Furthermore, in Examples 1,2, and 3, DIRT is compared with two other widely used methods to compute the probability of failure, namely, Bayesian Updating with Structural Reliability - Subset Simulation (BUS-SuS) and the Cross Entropy (CE) method. In Example 4, DIRT is only compared with BUS-SuS since CE becomes computationally expensive for the high-dimensional posterior failure probability example.   

\subsection{Linear Limit State Function}
This example is a proof of concept of our DIRT implementation. We show that the method works for a high-dimensional problem with small values of $\mathbb{P}(\mathcal{F})$ and consistently outperforms existing techniques by achieving a lower CoV.

Assume a limit state function given as a linear combination of $d$ independent standard Gaussian random variables
\begin{equation}
    g(\theta) = \alpha - \frac{1}{\sqrt{d}} \sum_{i=1}^d\theta_i
\end{equation}
where $\alpha$ is the threshold. It is used to control the value of $\mathbb{P}(\mathcal{F})$. For this specific limit state function, the analytical $\mathbb{P}(\mathcal{F})$ can be derived as $\mathbb{P}(\mathcal{F}) = \Phi(-\alpha)$ where $\Phi$ is the standard Gaussian Cumulative Distribution Function. Note that the analytical solution for $\mathbb{P}(\mathcal{F})$ is independent of the dimension $d$. We conduct a parametric study and choose the values of $\alpha = -\Phi^{-1}(\mathbb{P}(\mathcal{F}))$ from the set $[2.5,3.5,4.5,5.5,6.5,7.5]$. This corresponds to the target failure probabilities $\mathbb{P}(\mathcal{F}) = [6.210\times10^{-3},\ 2.326\times10^{-4},\ 3.398\times10^{-6},\ 1.900\times10^{-9},\ 4.016\times10^{-11},\ 3.200\times10^{-14}]$. For each threshold, we run the DIRT framework for $d = 2,25,50,75,100$. 

\begin{table}[htbp]
    \centering
    \resizebox{\textwidth}{!}{\begin{tabular}{c | c | c c | c c | c c}
        \toprule
        \( d \) & Ground Truth &
        \multicolumn{2}{c|}{DIRT} &
        \multicolumn{2}{c|}{SuS} &
        \multicolumn{2}{c}{CE} \\
        & \( \mathbb{P}(\mathcal{F}) = \Phi(-\alpha)\) & Mean $\Pfest$\ & CoV (\# evals) & Mean $\Pfest$\ & CoV (\# evals) & Mean $\Pfest$\ & CoV (\# evals)\\
        \midrule
        2   & \multirow{5}{*}{\( 2.326 \times 10^{-4} \)} & \( \mathbf{2.30 \times 10^{-4}} \) & \textbf{0.0229} (1440) & \( 2.52 \times 10^{-4} \) & 0.136 (15000) & \( 2.87 \times 10^{-4} \) & 0.36 (18000) \\
        25  & & \( \mathbf{2.30 \times 10^{-4}} \) & \textbf{0.0407} (17280) & \( 2.45 \times 10^{-4} \) & 0.158 (15000) & \( 2.19 \times 10^{-4} \) & 0.090 (150000) \\
        50  & & \( 2.36 \times 10^{-4} \) & \textbf{0.0258} (45280) & \( \mathbf{2.32 \times 10^{-4}} \) & 0.129 (15000) & \( 1.76 \times 10^{-4} \) & 0.555 (500000) \\
        75  & & \( \mathbf{2.32 \times 10^{-4}} \) & \textbf{0.0086} (69120) & \( 2.29 \times 10^{-4} \) & 0.145 (15000) & \( 2.27 \times 10^{-4} \) & 0.118 (900000) \\
        100 & & \( \mathbf{2.33 \times 10^{-4}} \) & \textbf{0.0235} (81820) & \( 2.28 \times 10^{-4} \) & 0.206 (15000) & \( 1.22 \times 10^{-4} \) & 1.104 (1200000) \\
        \bottomrule
    \end{tabular}}
    \vspace{10pt}
    \caption{Comparison of DIRT, SuS, and Cross-Entropy (CE) methods for estimating $\Pfest$ at \( \alpha = 3.5 \) across increasing dimension \( d \). The ground truth $\Pfest$ is \( \Phi(-\alpha) \). Best (closest to ground truth) $\Pfest$ and lowest CoV in each row are shown in bold.}
    \label{tab:dirt_sus_ce_beta3p5}
\end{table}

We compare the performance of DIRT with standard Subset Simulation (SuS) to estimate \( \mathbb{P}(\mathcal{F}) \). The comparison focuses on two key metrics: the coefficient of variation (CoV) of the estimated \( \mathbb{P}(\mathcal{F}) \), and the computational cost measured as the number of limit state function evaluations. We keep the maximum TT rank of the density approximations as $r = 2$ for $d= \{2,25\}$ and $r = 4$ for $d = {50,75,100}$, and we set the number of tempering levels  $L = 12$ for all dimensions. For BUS-SuS, we set the number of samples per level $N = 3000$ and the probability of each level $p = 0.1$, following the parameter choices established in \cite{Subset_original}. For CE, the number of samples per level was kept as $N = \{3000,10000,20000,50000,100000\}$ for $d = \{2,25,50,75,100\}$ respectively.   

\begin{figure}
    \centering
    \includegraphics[width=0.90\linewidth]{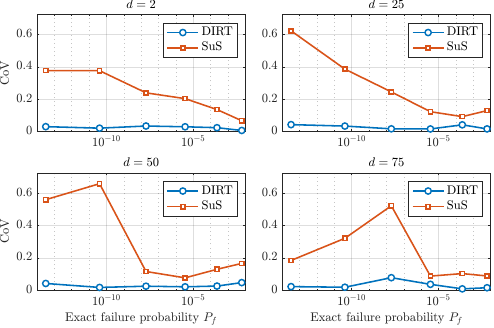}
    \caption{Coefficient of Variation (CoV) of \( \mathbb{P}(\mathcal{F}) \) estimates achieved by DIRT as a function of exact \( \mathbb{P}(\mathcal{F}) \)}
    \label{fig:cov_vs_pf}
\end{figure}

Table~\ref{tab:dirt_sus_ce_beta3p5} summarizes the estimated mean \( \mathbb{P}(\mathcal{F}) \), CoV, relative error with respect to the analytical \( \mathbb{P}(\mathcal{F}) \), and the number of function evaluations for all methods across all reliability indices \( \alpha = 2.5,3.5,4.5,5.5,6.5,7.5 \). The DIRT method consistently achieves a much lower CoV than SuS and CE, often by an order of magnitude, indicating more stable estimates even in high-dimensional or rare-event scenarios. Moreover, the accuracy of the \( \mathbb{P}(\mathcal{F}) \) estimates from DIRT remains high, with relative errors below \( 1.5\% \) in all cases. Table~\ref{tab:dirt_sus_ce_d100} compares the mean and CoV of $\Pfest$ across different values of $\alpha$ and $d = 100$. In terms of computational cost, the number of function evaluations with DIRT depends primarily on the problem dimension \( d \), but remains nearly independent of the analytical \( \mathbb{P(\mathcal{F})} \). In contrast, SuS requires substantially more evaluations as \( \alpha \) decreases, due to the need to sample additional intermediate subsets for rare events. Lastly, the number of evaluations required in CE blows up with $d$ while also leading to poorer approximations of $\mathbb{P(\mathcal{F})}$. It was observed that for $d = 100$ and $\alpha \geq4.5$, CE was unable to estimate the correct $\Pf$ even for $500,000$ samples per level. This makes DIRT particularly advantageous in rare-event settings. 

\begin{table}[htbp]
    \centering
    \resizebox{\textwidth}{!}{\begin{tabular}{c | c | c c | c c | c c}
        \toprule
        \( \alpha \) & Ground Truth &
        \multicolumn{2}{c|}{DIRT} &
        \multicolumn{2}{c|}{SuS} &
        \multicolumn{2}{c}{CE} \\
        & \( \mathbb{P}(\mathcal{F}) = \Phi(-\alpha)\) & Mean $\Pfest$\ & CoV (\# evals) & Mean $\Pfest$\ & CoV (\# evals) & Mean $\Pfest$\ & CoV (\# evals)\\
        \midrule
        2.5   & \( 6.210\times10^{-3} \) & \( 6.24 \times 10^{-3} \) & 0.021 (81280) & \( \mathbf{6.20 \times 10^{-3}} \) & 0.049 (15000) & \( 6.19 \times 10^{-3} \) & \textbf{0.005} (800000) \\
        3.5  &\( 2.326 \times 10^{-4} \) & \( \mathbf{2.33 \times 10^{-4}} \) & \textbf{0.023} (81280) & \( 2.28 \times 10^{-4} \) & 0.205 (21000) & \( 1.22 \times 10^{-4} \) & 1.104 (1200000) \\
        4.5  &\( 3.398\times10^{-6} \) & \( \mathbf{3.41 \times 10^{-6}} \) & \textbf{0.002} (81280) & \( 3.46 \times 10^{-6} \) & 0.291 (27000) & $-$ & $-$ \\
        5.5  &\( 1.900\times10^{-9} \) & \( \mathbf{1.88 \times 10^{-8}} \) & \textbf{0.011} (81280) & \( 2.10 \times 10^{-8} \) & 0.277 (36000) & $-$ & $-$ \\
        6.5  &\( 4.016\times10^{-11} \) & \( \mathbf{3.94 \times 10^{-11}} \) & \textbf{0.067} (81280) & \( 5.27 \times 10^{-11} \) & 0.486 (45000) & $-$ & $-$ \\
        7.5 &\( 3.200\times10^{-14} \) & \( \mathbf{3.11 \times 10^{-14}} \) & \textbf{0.019} (81820) & \( 2.66 \times 10^{-14} \) & 0.451 (54000) & $-$ & $-$ \\
        \bottomrule
    \end{tabular}}
    \vspace{10pt}
    \caption{Comparison of DIRT, SuS, and Cross-Entropy (CE) methods for estimating $\Pfest$ at \( d=100 \) across increasing \( \alpha \). The ground truth $\Pfest$ is \( \Phi(-\alpha) \). Best (closest to ground truth) $\Pfest$ and lowest CoV in each row are shown in bold.}
    \label{tab:dirt_sus_ce_d100}
\end{table}

Figure~\ref{fig:cov_vs_pf} plots the CoV of both methods as a function of $\Pfest$. The figure highlights the stability of DIRT across dimensions and failure probabilities, while SuS exhibits larger and more variable CoV values as the true $\Pfest$ gets smaller.  While DIRT’s computational cost grows with dimension, the method remains more efficient and reliable than SuS, particularly in the rare-event regime.   

\subsection{4 Dimensional Posterior Example - Corroded Beam}
\begin{figure}
    \centering
    \includegraphics[width=\linewidth]{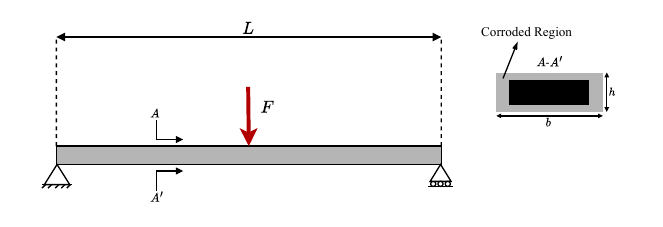}
    \caption{Corroded beam subjected to midspan load}
    \label{fig:beam}
\end{figure}

This numerical example investigates the behavior of an Euler-beam subjected to a load $F$ as shown in Figure \ref{fig:beam}.
The limit state function is given by
\begin{equation}
    g(h,b) = \sigma(h,b) - \sigma_{\text{max}}
\end{equation}
where $\sigma_{\max} = 500$ MPa is the maximum allowable stress. $\sigma(h,b)$ is the maximum normal stress across the beam, and it can be derived as follows
\begin{equation}
    \sigma(h,b) = \frac{M(h,b)}{W(h,b)},
\end{equation}
M and W denote the maximum bending moment and bending section coefficient, respectively, derived as
\begin{equation}
\begin{aligned}
        W = \frac{b h^2}{6}, \quad 
M = \frac{F L}{4} + \frac{\rho_{\text{st}} b h L^2}{8},
\end{aligned}
\end{equation}
where $\rho_{st}$ is the density of steel, $L$ is the length of the beam, $b$ is the width of the beam section, and $h$ is the height of the beam section. 

We model $b$, $h$, $F$, and $L$ as random variables. The cross-sectional dimensions $b$ and $h$ are assumed to follow normal distributions, with their respective means ($\mu_b$, $\mu_h$) and standard deviations ($\sigma_b$, $\sigma_h$) treated as uncertain quantities. The four parameters $\mu_b$, $\sigma_b$, $\mu_h$, and $\sigma_h$ are considered hyperparameters whose posterior distributions are inferred from the data. The applied force $F$ and the length $L$ are modeled as log-normally distributed random variables with deterministic means and standard deviations. In addition, a correlation coefficient of 0.4 is assumed between $b$ and $h$. A summary of the prior distributions and their associated statistical parameters is provided in Table~\ref{table:priors}.

\begin{table}[h!]
\centering

\begin{tabular}{cccc}
\hline
\vspace{2.5pt}
\textbf{Variable} & \textbf{Distribution} & \textbf{Mean} & \textbf{Standard deviation}\\ \hline
$b$ (m) & Normal & $N(0.2, 0.03)$ & LN$(0.03, 4.5 \times 10^{-3})$\\ 
$h$ (m) & Normal & $N(0.03, 4.5 \times 10^{-3})$ & LN$(4.5 \times 10^{-3}, 6.75 \times 10^{-4})$ \\
$F$ (N) & Lognormal & $3500$ & $700$  \\ 
$L$ (m) & Lognormal & $5$ & $0.5$  \\ 
$\rho_{\text{st}}$ (kN$\cdot$m$^{-3}$) & Deterministic & $78.5$ & $-$  \\ \hline
\end{tabular}
\vspace{10pt}
\caption{Prior Distributions for Example 2}
\label{table:priors}
\end{table}

We assume observational data for $b$ and $h$ as the beam is corroded over time as $D_{X,1} = \{D_{b,1} = 0.18, \, D_{h,1} = 0.026\}$ and $D_{X,2} = \{D_{b,2} = 0.14, \, D_{h,2} = 0.019\}$. The likelihood function is then given as 
\begin{equation}
\prod_{i=1}^{N_d} \frac{1}{\sqrt{(2\pi)^{N_d} |\Sigma_0|}} 
\exp \left( -\frac{1}{2} (D_{X,i} - \mu_0)^\top \Sigma_0^{-1} (D_{X,i} - \mu_0) \right),
\end{equation}
where the mean vector and the covariance are given by $\mu_0 = \begin{pmatrix} \mu_b \\ \mu_h \end{pmatrix}^\top$ and  $\Sigma_0 = 
\begin{pmatrix}
\sigma_b^2 & 0.4 \sigma_b \sigma_h \\
0.4 \sigma_b \sigma_h & \sigma_h^2
\end{pmatrix}$.

\begin{figure}
    \centering
    \includegraphics[width=0.85\linewidth]{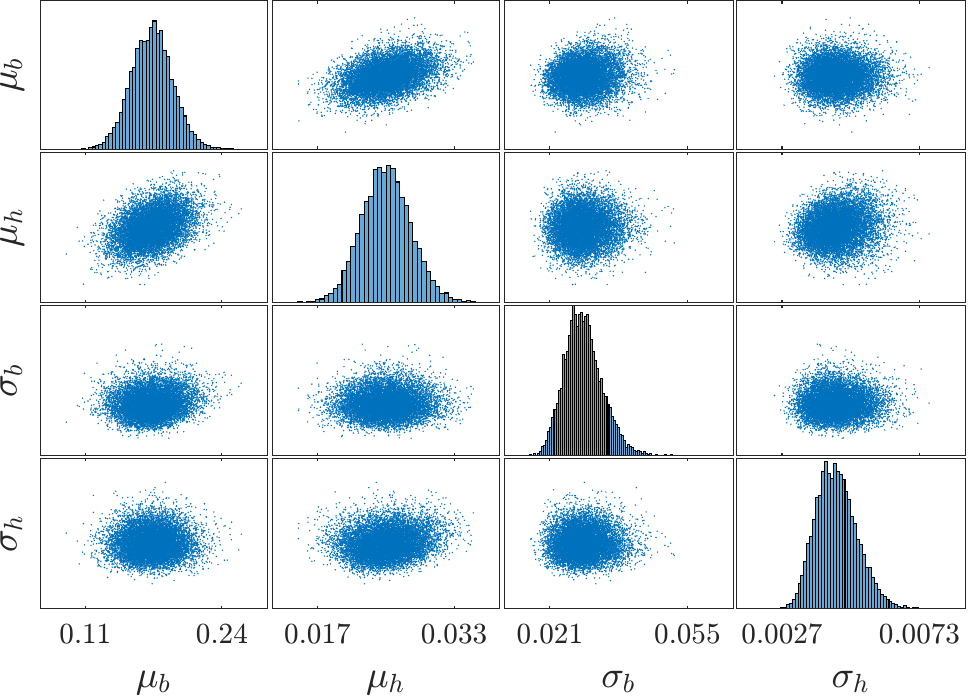}
    \caption{Joint posterior samples of the unknown parameters in Example 2. Diagonal: marginal distributions. Off-diagonal: pairwise joint distributions.}
    \label{fig:posterior_corrodedBeam}
\end{figure}

We compare the obtained probability of failure between DIRT, BUS-SuS, and CE. For the DIRT framework, we fixed the maximum TT approximation rank to $r = 2$ since this is a low-dimensional example. The number of bridging measures $L$, was kept as 4 by tempering the posterior density as $\pi_k(x) = \pi(x)^{\beta_k}$, starting with $\beta_0 = 10^{-4}$ and the $(k+1)^{th}$ bridging measure is given by $\beta_{k+1} = 10.\beta_{k}$. A Gaussian distribution with mean 0 and standard deviation 3 is chosen as the reference distribution $\rho$. The input parameters $[\mu_b, \mu_h, \sigma_b, \sigma_h]$ are scaled to be in the domain $(0.5, 1.5)$, restricting them to the interval $[a_k, b_k]$, where $a = [0.1, 0.015, 0.015, 0.00225]$ and $b = [0.3, 0.045, 0.045, 0.00675]$. 
For BUS-SuS, again we use 3000 samples per level and the probability of each subset is set as $p = 0.1$ following Ref. \cite{Subset_original}. 
\begin{table}[h!]
\centering
\begin{tabular}{cccccc}
\toprule
Update & Method & $\Pfpostest$ & $\hat{\mathbb{P}}_{0.75}(\mathcal{F} \mid \tilde{\mathbf{y}})$ & $\hat{\mathbb{P}}_{0.9}(\mathcal{F} \mid \tilde{\mathbf{y}})$ \\
\midrule
\multirow{3}{*}{1\textsuperscript{st}} 
& BUS-SuS         & $7.296 \times 10^{-2}$ & $9.999 \times 10^{-2}$ & $1.702 \times 10^{-1}$ \\
& CE   & $6.800 \times 10^{-2}$ & $9.500 \times 10^{-2}$ & $1.650 \times 10^{-1}$ \\
& DIRT            & $7.021 \times 10^{-2}$ & $1.000 \times 10^{-1}$ & $1.700 \times 10^{-1}$ \\
\midrule
\multirow{3}{*}{2\textsuperscript{nd}} 
& BUS-SuS         & $1.636 \times 10^{-1}$ & $2.209 \times 10^{-1}$ & $3.126 \times 10^{-1}$ \\
& CE   & $1.580 \times 10^{-1}$ & $2.150 \times 10^{-1}$ & $3.000 \times 10^{-1}$ \\
& DIRT            & $1.636 \times 10^{-1}$ & $2.300 \times 10^{-1}$ & $3.300 \times 10^{-1}$ \\
\bottomrule
\end{tabular}
\vspace{10pt}
\caption{Comparison of $\Pfpostest$ between BUS-SuS, Cross Entropy, and DIRT for Example 2.}
\label{tab:pf_cb}
\end{table}

\begin{figure}[htbp]
    \centering
    \begin{subfigure}{0.45\linewidth}
        \centering
        \includegraphics[height=5cm]{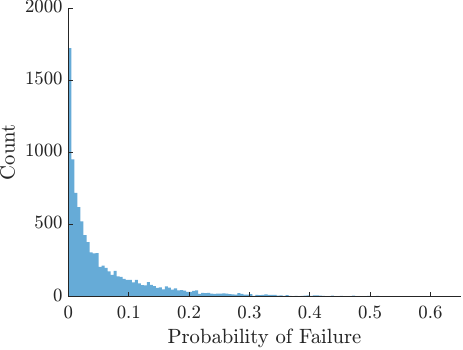}
        \caption{After 1st update}
        \label{fig:pf1_cb}
    \end{subfigure}
    \hfill
    \begin{subfigure}{0.45\linewidth}
        \centering
        \includegraphics[height=5cm]{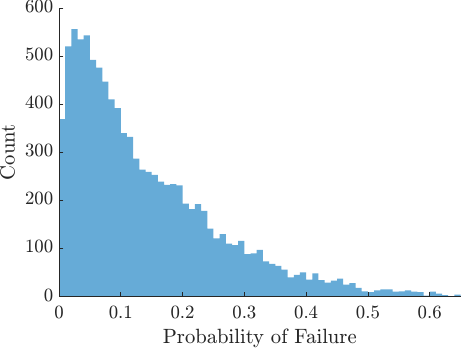}
        \caption{After 2nd Update}
        \label{fig:pf2_cb}
    \end{subfigure}
    \caption{Distribution of the posterior probability of failure in Example 2}
    \label{fig:pf_cb}
\end{figure}

\vspace{0.5em}

As shown in Table~\ref{tab:pf_cb}, the posterior failure probability estimates obtained using DIRT closely match those from BUS-SuS across both updates and at multiple quantile levels. After the first update, the failure probabilities predicted by DIRT and BUS-SuS are \( 7.021 \times 10^{-2} \) and \( 7.296 \times 10^{-2} \), respectively, with a relative error of less than 4\%. Similarly, after the second update, all three methods report a mean value of around \( 1.6 \times 10^{-1} \), with minor variation in the upper quantiles.

\vspace{0.5em}

This consistency confirms that the DIRT framework is able to reliably approximate the tail of the posterior distribution governing failure. The agreement across quantiles \( P_{f,0.75} \) and \( P_{f,0.9} \) further illustrates DIRT’s robustness.

\vspace{0.5em}

Figures~\ref{fig:pf1_cb} and \ref{fig:pf2_cb} show the distribution of the estimated probability of failure across posterior samples.

\subsection{Cantilever Beam with Variable Flexibility}
\begin{figure}
    \centering
    \includegraphics[width=0.65\linewidth]{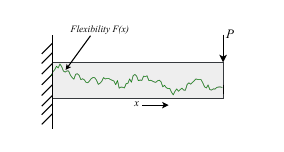}
    \caption{Cantilever beam with variable flexibility subjected to load $P$}
    \label{fig:cantileverbeam}
\end{figure}
This example consists of a cantilever beam of length 2 m with a point load P = 20 kN applied at the free end. The spatially varying flexibility of the beam, F(x), is equal to (1 / E(x)I), where E is the Young's modulus and I is the moment of inertia. The mean flexibility is $\mu_F = 1\times10^{-4}kN/m2$ and the standard deviation is $\sigma_F = 3.5\times10^{-5}kN/m^2$. The true flexibility is modeled using one realization of a 1-D Gaussian random field with an exponential kernel with correlation length $l_{c} = 2.0$.  Observational data of the beam deflection w is generated using the true flexibility field as shown in Figure 1. The deflections are measured at equally spaced points along the beam and contain some noise $\eta$. Noise is modeled using an exponential kernel with standard deviation $\sigma_{\eta} = 1\times10^{-3}$ and correlation length $l_{\eta} = 1.0$. This makes the noise Gaussian with 0 mean and covariance matrix $\Sigma_{\eta\eta}$. This leads to the following likelihood function

\begin{equation}
\mathcal{L}(F(x); \tilde{\mathbf{y}}) = \frac{1}{\sqrt{(2\pi)^m \det(\boldsymbol{\Sigma}_{\eta\eta})}} 
\exp\left(-\frac{1}{2}[\tilde{\mathbf{y}} - w(\tilde{\mathbf{x}}, F(x))]^{\top} 
\boldsymbol{\Sigma}_{\eta\eta}^{-1}[\tilde{\mathbf{y}} - w(\tilde{\mathbf{x}}, F(x))] \right),
\end{equation}
here $F(x)$ is a realization of the flexibility random field, and 
$\tilde{\mathbf{y}}$ is a set of $m$ deflection observations.

\begin{figure}[htbp]
    \centering
    \begin{subfigure}[t]{0.45\textwidth}
        \centering
        \includegraphics[width=\linewidth]{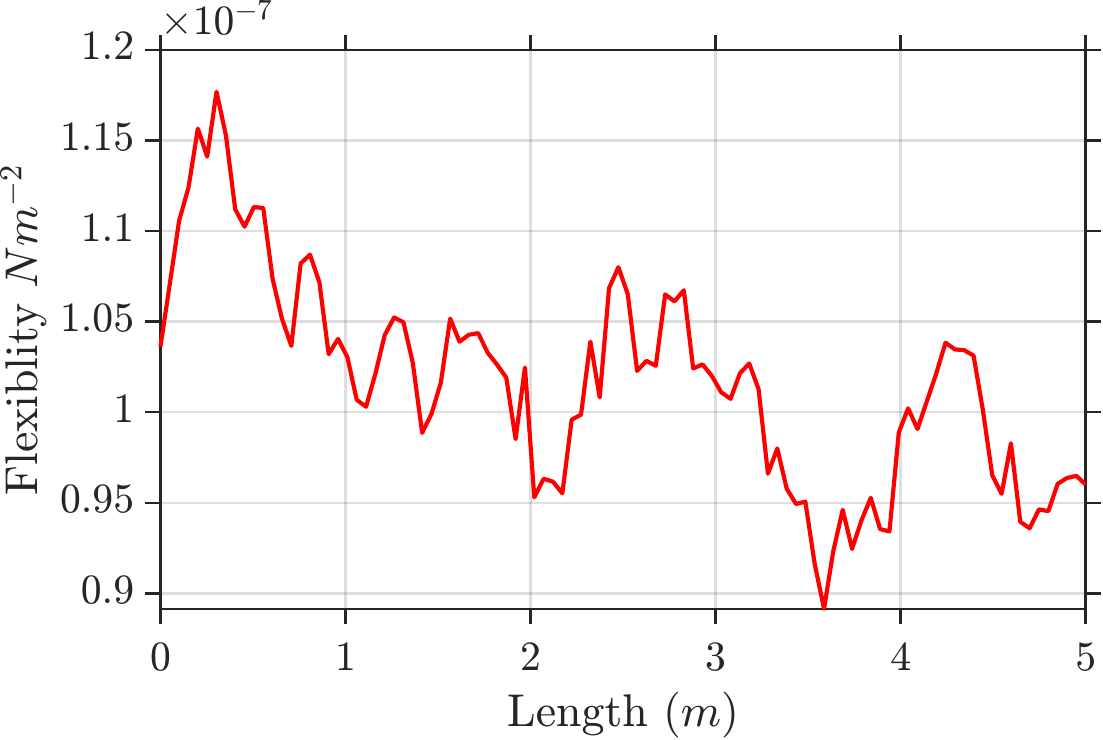} 
        \caption{True Flexibility}
    \end{subfigure}
    \hfill
    \begin{subfigure}[t]{0.45\textwidth}
        \centering
        \includegraphics[width=\linewidth]{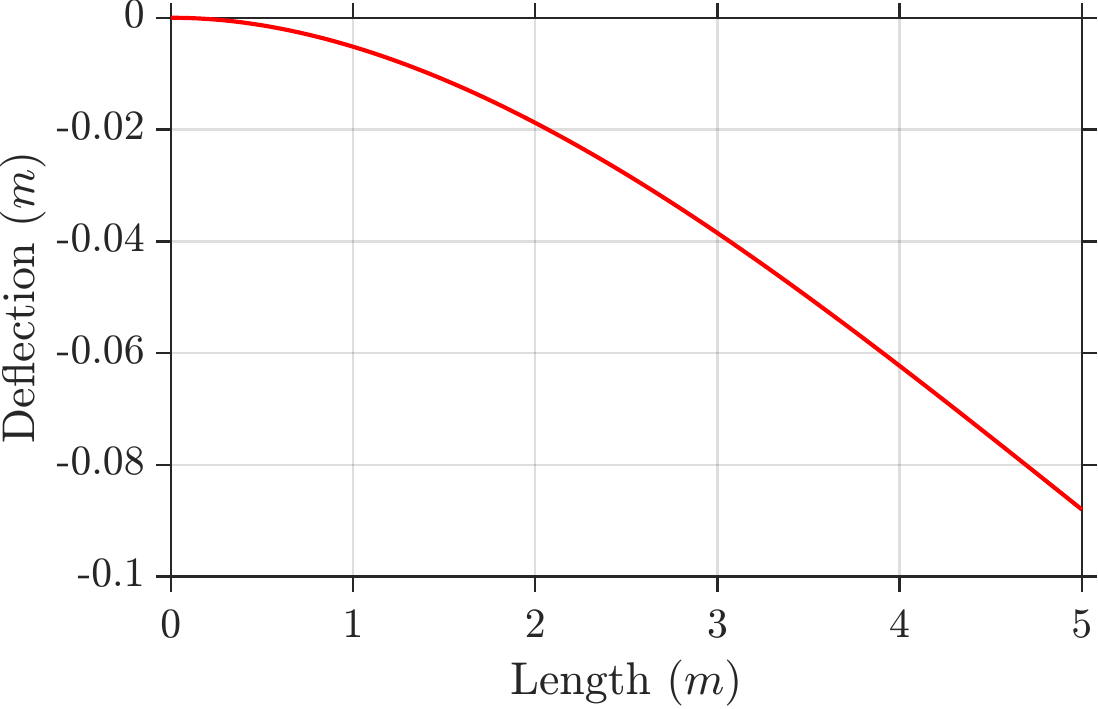}
        \caption{True Deflection}
    \end{subfigure}

    \vspace{0.45cm}

    \begin{subfigure}[t]{0.45\textwidth}
        \centering
        \includegraphics[width=\linewidth]{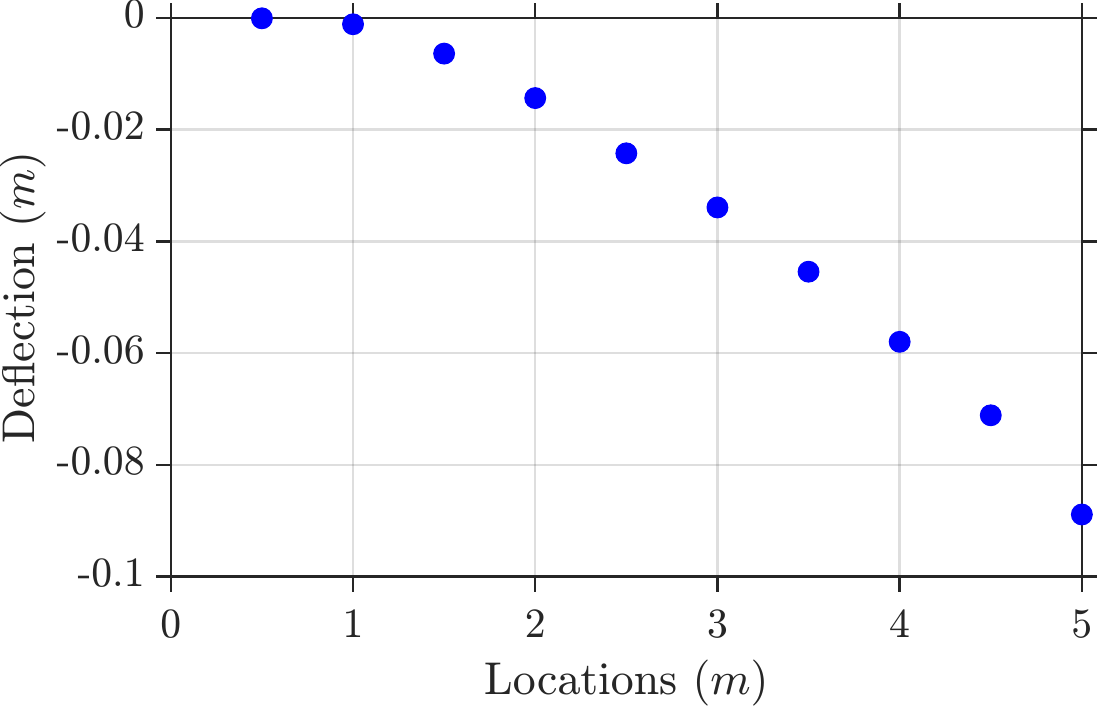}
        \caption{Observed noisy data for m = 10}
    \end{subfigure}
    \hfill
    \begin{subfigure}[t]{0.45\textwidth}
        \centering
        \includegraphics[width=\linewidth]{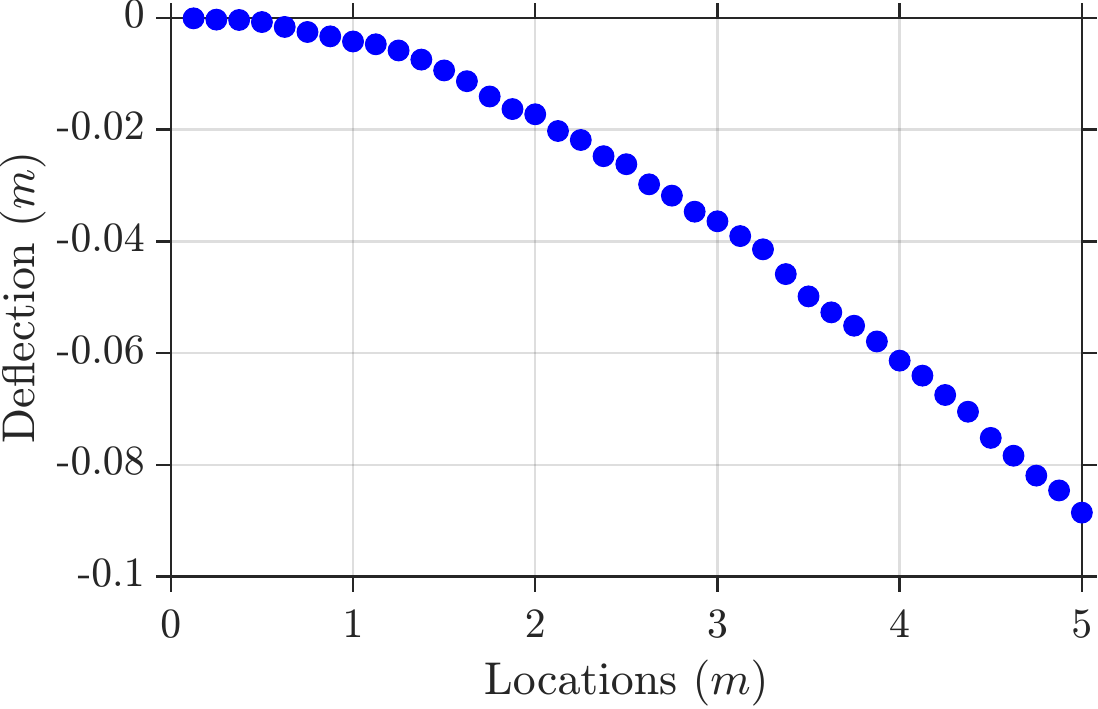}
        \caption{Observed noisy data for m = 40}
    \end{subfigure}

    \caption{Cantilever beam: true values and two sets of deflection observations.}
\end{figure}

\newcommand{\rowlabel}[1]{%
  \raisebox{-0.5\height}{\rotatebox[origin=c]{90}{\large\(\boldsymbol{#1}\)}}}
\begin{figure}[ht]
  \centering
  \setlength{\tabcolsep}{3pt}
  \renewcommand{\arraystretch}{0.9}

  \begin{tabular}{@{}r *{3}{>{\centering\arraybackslash}m{0.29\textwidth}}@{}}
       & \textbf{$\ell_c = 0.5$} & \textbf{$\ell_c = 2.5$} & \textbf{$\ell_c = 4.5$} \\[2pt]

    \rowlabel{m = 10} &
      \begin{subfigure}[t]{\linewidth}\includegraphics[width=\linewidth]{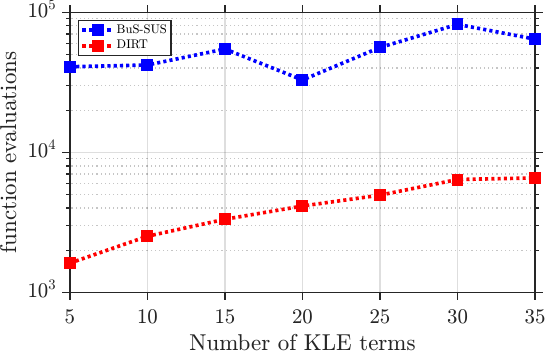}\end{subfigure} &
      \begin{subfigure}[t]{\linewidth}\includegraphics[width=\linewidth]{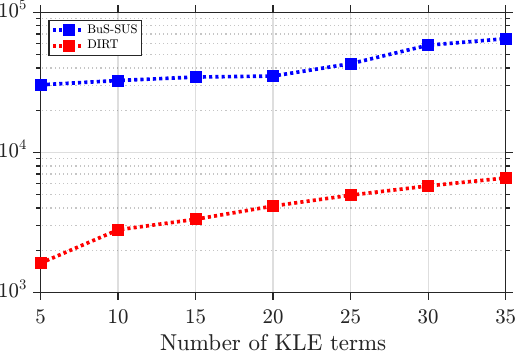}\end{subfigure} &
      \begin{subfigure}[t]{\linewidth}\includegraphics[width=\linewidth]{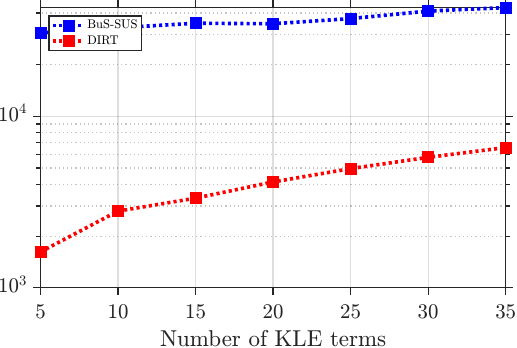}\end{subfigure} \\[6pt]

    \rowlabel{m = 40} &
      \begin{subfigure}[t]{\linewidth}\includegraphics[width=\linewidth]{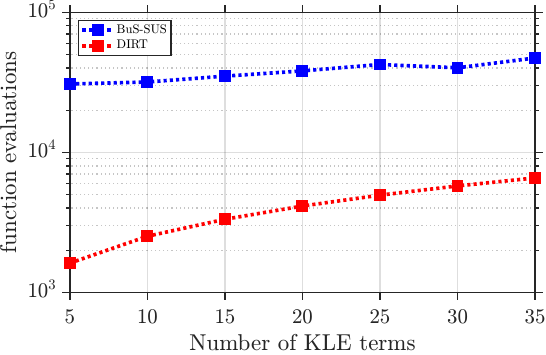}\end{subfigure} &
      \begin{subfigure}[t]{\linewidth}\includegraphics[width=\linewidth]{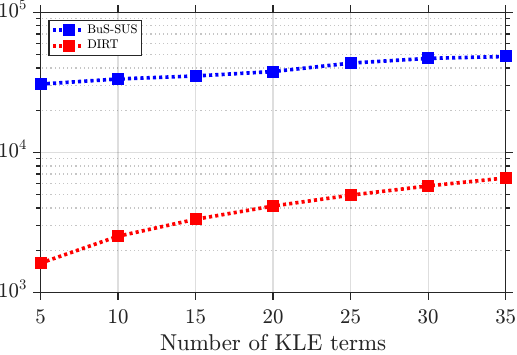}\end{subfigure} &
      \begin{subfigure}[t]{\linewidth}\includegraphics[width=\linewidth]{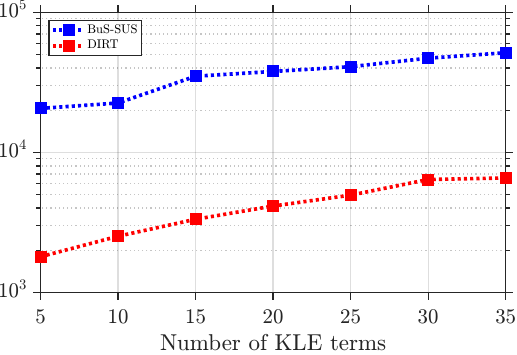}\end{subfigure} \\
  \end{tabular}

  \caption{Function evaluations vs.\ \(M\) for the six \((m,\ell_c)\) combinations.}
  \label{fig:nevals_grid}
\end{figure}

To represent the prior flexibility field \( F(x) \) in a reduced-dimensional form, we employ the \textit{Karhunen--Lo\`eve (KL) expansion}. The KL expansion provides an efficient way to express a second-order stochastic process in terms of an orthogonal basis derived from its covariance structure. Given a mean function \( \mu_F(x) = \mathbb{E}[F(x)] \) and a covariance function \( C_F(x_1, x_2) = \mathrm{Cov}[F(x_1), F(x_2)] \), the field can be approximated as:
\begin{equation}
F(x, \boldsymbol{\xi}) \approx \mu_F(x) + \sum_{i=1}^{M} \sqrt{\lambda_i} \, \phi_i(x) \, \xi_i,
\end{equation}
where:
\begin{itemize}
    \item \( \{ \lambda_i, \phi_i(x) \} \) are the eigenpairs of the covariance kernel \( C_F \), satisfying the Fredholm integral equation:
    \begin{equation}
    \int C_F(x, x') \phi_i(x') \, dx' = \lambda_i \phi_i(x),
    \end{equation}
    \item \( \xi_i \sim \mathcal{N}(0, 1) \) are uncorrelated standard normal random variables,
    \item \( M \) is the number of retained modes (truncation parameter).
\end{itemize}

This gives us M random variables, the posterior distribution of which can be inferred using the DIRT framework. 

The number of tempering layers is fixed as $L = 4$, and the maximum TT rank is fixed as $4$. The smoothing parameter $\gamma$ is informed by exploratory testing and set to $1000$. In BUS-SuS, the number of samples per level is set to 10000 following Ref. \cite{BUS_varying}, and the probability of each subset is set to 0.1; in CE, the number of samples per level is set to $2^{14}$ following Ref. \cite{DIRT2021}.

We vary the following parameters of the KL expansion to check their influence on the random fields

\begin{itemize}
    \item Number of modes in the KL expansion, M = {5,10,50}
    \item Correlation length of the prior, $l_c$ = {0.45,2.5,4.5} m
    \item number of measurement data locations m = {10,40}
\end{itemize}

The beam is considered to fail when the tip deflection $\delta(\mathbf{x})$ exceeds a threshold $\delta_{max}$. This leads to the following limit-state function
\begin{equation}
    g(\mathbf{x}) = \delta_{max} - \delta(\mathbf{x}).
\end{equation}
For the rare event, we choose $\delta_{max} = L / 55$. 
    
\begin{table}[t!]
    \centering
    \renewcommand{\arraystretch}{1.2}
    \begin{tabular}{c c c c c c c}
        \multicolumn{7}{c}{$\ell_c = \mathbf{0.5}$} \\
        \toprule
        \multicolumn{1}{c}{$M$} &
        \multicolumn{2}{c}{\textbf{BUS-SuS}} &
        \multicolumn{2}{c}{\textbf{Cross Entropy}} &
        \multicolumn{2}{c}{\textbf{DIRT}} \\
        & $\mathbb{P}[\mathcal{F} \mid \tilde{y}]$ & CoV & $\mathbb{P}[\mathcal{F} \mid \tilde{y}]$ & CoV & $\mathbb{P}[\mathcal{F} \mid \tilde{y}]$ & CoV \\
        \midrule
         5   & $2.20 \times 10^{-6}$ & $0.747$ & $4.28 \times 10^{-6}$ & $\mathbf{8.87 \times 10^{-3}}$ & $4.10 \times 10^{-6}$ & $1.38\times10^{-2}$ \\
        10   & $4.92 \times 10^{-6}$ & $0.272$ & $1.86 \times 10^{-6}$ & $1.20$ & $4.13 \times 10^{-6}$ & $7\mathbf{.65\times10^{-3}}$ \\
        20   & $3.08 \times 10^{-6}$ & $0.340$ & $3.36 \times 10^{-6}$ & $2.07$ & $4.11 \times 10^{-6}$ & $\mathbf{6.74\times10^{-3}}$ \\
        
        \midrule

        \multicolumn{7}{c}{$\ell_c = \mathbf{2.5}$} \\
        \toprule
        $M$ & $\mathbb{P}[\mathcal{F} \mid \tilde{y}]$ & CoV & $\mathbb{P}[\mathcal{F} \mid \tilde{y}]$ & CoV & $\mathbb{P}[\mathcal{F} \mid \tilde{y}]$ & CoV \\
        \midrule
         5   & $2.58 \times 10^{-5}$ & $0.255$ & $2.46 \times 10^{-5}$ & $\mathbf{4.90\times10^{-3}}$ & $2.37 \times 10^{-5}$ & $6.40\times10^{-3}$ \\
        10   & $3.28 \times 10^{-5}$ & $0.0773$ & $3.73 \times 10^{-5}$ & $7.74\times10^{-1}$ & $2.37 \times 10^{-5}$ & $\mathbf{1.33\times10^{-2}}$ \\
        20   & $2.44 \times 10^{-5}$ & $0.390$ & $3.94 \times 10^{-5}$ & $5.46\times10^{-1}$ & $2.36 \times 10^{-5}$ & $\mathbf{6.93\times10^{-3}}$ \\
        
        \midrule

        \multicolumn{7}{c}{$\ell_c = \mathbf{4.5}$} \\
        \toprule
        $M$ & $\mathbb{P}[\mathcal{F} \mid \tilde{y}]$ & CoV & $\mathbb{P}[\mathcal{F} \mid \tilde{y}]$ & CoV & $\mathbb{P}[\mathcal{F} \mid \tilde{y}]$ & CoV \\
        \midrule
         5   & $3.70 \times 10^{-5}$ & $0.144$ & $3.08 \times 10^{-5}$ & $7.15\times10^{-3}$ & $2.97 \times 10^{-5}$ & $\mathbf{7.34\times10^{-3}}$ \\
        10   & $3.13 \times 10^{-5}$ & $0.308$ & $2.58 \times 10^{-5}$ & $4.30\times10^{-1}$ & $2.99 \times 10^{-5}$ & $\mathbf{1.20\times10^{-2}}$ \\
        20   & $2.37 \times 10^{-5}$ & $0.374$ & $5.30 \times 10^{-5}$ & $1.87\times10^{-1}$ & $2.98 \times 10^{-5}$ & $\mathbf{5.07\times10^{-3}}$ \\
        
        \bottomrule
    \end{tabular}
    \vspace{20pt}
    \caption{Summary and comparison of posterior failure probabilities between BUS-SuS, Cross Entropy method, and DIRT for m = 10}
    \label{tab:beam_results1}
\end{table}
\begin{table}[t!]
    \centering
    \renewcommand{\arraystretch}{1.2}
    \begin{tabular}{c c c c c c c}
        \multicolumn{7}{c}{$\ell_c = \mathbf{0.5}$} \\
        \toprule
        \multicolumn{1}{c}{$M$} &
        \multicolumn{2}{c}{\textbf{BUS-SuS}} &
        \multicolumn{2}{c}{\textbf{Cross Entropy}} &
        \multicolumn{2}{c}{\textbf{DIRT}} \\
        & $\mathbb{P}[\mathcal{F} \mid \tilde{y}]$ & CoV & $\mathbb{P}[\mathcal{F} \mid \tilde{y}]$ & CoV & $\mathbb{P}[\mathcal{F} \mid \tilde{y}]$ & CoV \\
        \midrule
         5   & $2.38 \times 10^{-5}$ & $4.21 \times 10^{-1}$ & $2.05 \times 10^{-5}$ & $1.21 \times 10^{-2}$ & $2.00 \times 10^{-5}$ & $\mathbf{7.21 \times 10^{-3}}$ \\
        10   & $1.81 \times 10^{-5}$ & $1.97 \times 10^{-1}$ & $1.31 \times 10^{-5}$ & $7.80 \times 10^{-1}$ & $1.99 \times 10^{-5}$ & $\mathbf{7.75 \times 10^{-3}}$ \\
        20   & $2.97 \times 10^{-5}$ & $3.66 \times 10^{-1}$ & $4.79 \times 10^{-6}$ & $1.41 \times 10^{+0}$ & $2.01 \times 10^{-5}$ & $\mathbf{9.38 \times 10^{-3}}$ \\
        
        \midrule

        \multicolumn{7}{c}{$\ell_c = \mathbf{2.5}$} \\
        \toprule
        $M$ & $\mathbb{P}[\mathcal{F} \mid \tilde{y}]$ & CoV & $\mathbb{P}[\mathcal{F} \mid \tilde{y}]$ & CoV & $\mathbb{P}[\mathcal{F} \mid \tilde{y}]$ & CoV \\
        \midrule
         5   & $1.07 \times 10^{-4}$ & $1.76 \times 10^{-1}$ & $1.06 \times 10^{-4}$ & $\mathbf{2.65 \times 10^{-3}}$ & $1.02 \times 10^{-4}$ & $6.95 \times 10^{-3}$ \\
        10   & $9.60 \times 10^{-5}$ & $1.85 \times 10^{-1}$ & $1.62 \times 10^{-4}$ & $7.80 \times 10^{-1}$ & $1.20 \times 10^{-4}$ & $\mathbf{1.51 \times 10^{-2}}$ \\
        20   & $9.01 \times 10^{-5}$ & $3.38 \times 10^{-1}$ & $1.77 \times 10^{-4}$ & $5.55 \times 10^{-1}$ & $1.02 \times 10^{-4}$ & $\mathbf{5.24 \times 10^{-3}}$ \\
        
        \midrule

        \multicolumn{7}{c}{$\ell_c = \mathbf{4.5}$} \\
        \toprule
        $M$ & $\mathbb{P}[\mathcal{F} \mid \tilde{y}]$ & CoV & $\mathbb{P}[\mathcal{F} \mid \tilde{y}]$ & CoV & $\mathbb{P}[\mathcal{F} \mid \tilde{y}]$ & CoV \\
        \midrule
         5   & $1.28 \times 10^{-4}$ & $2.25 \times 10^{-1}$ & $1.27 \times 10^{-4}$ & $6.85 \times 10^{-3}$ & $1.24 \times 10^{-4}$ & $\mathbf{4.56 \times 10^{-3}}$ \\
        10   & $1.33 \times 10^{-4}$ & $1.73 \times 10^{-1}$ & $1.08 \times 10^{-4}$ & $3.67 \times 10^{-1}$ & $1.23 \times 10^{-4}$ & $\mathbf{1.24 \times 10^{-2}}$ \\
        20   & $1.37 \times 10^{-4}$ & $5.26 \times 10^{-1}$ & $3.55 \times 10^{-4}$ & $4.72 \times 10^{-1}$ & $1.22 \times 10^{-4}$ & $\mathbf{1.10 \times 10^{-2}}$ \\
        
        \bottomrule
    \end{tabular}
    \vspace{20pt}
    \caption{Summary and comparison of posterior failure probabilities between BUS-SuS, Cross Entropy method, and DIRT for $m = 40$.}
    \label{tab:beam_results2}
\end{table}

Figure~\ref{fig:nevals_grid} shows the number of function evaluations as a function of \(M\) for all combinations of \(\ell_c\) and \(m\). We observe that for all configurations, DIRT requires significantly fewer function evaluations compared to BUS-SuS, and its cost scales linearly as \(M\) increases. This shows the scalability of the DIRT framework with respect to the dimensionality of the KL representation.

The posterior failure probabilities and corresponding CoVs are summarized in Tables~\ref{tab:beam_results1} and~\ref{tab:beam_results2}. Again, across all configurations, the estimated probabilities of failure are comparable among all three methods. However, the CoV of the DIRT estimates is consistently lower than those of BUS-SuS and CE, indicating higher estimator reliability for DIRT.

We note that for this example, increasing the number of KL modes \(M\) has little impact on the estimated $\Pfpostest$, suggesting that the truncation at \(M=10\) already captures most of the variability in the flexibility field for the chosen correlation lengths. Increasing \(m\) also leads to tighter estimates, as expected.

\subsection{High Dimensional Finite Element Example - 2D Plate}
For the high-dimensional posterior example, we consider a square low-carbon steel plate with side lengths $0.32$ m with a hole of radius $0.02$ m at the center. The thickness of the plate is $0.01$ m. The material of the plate is assumed to vary spatially; specifically, the Young's Modulus is a function of the spatial coordinates. The left edge of the plate is subjected to the Dirichlet boundary condition $\mathbf{u(x) = 0}$.  The right edge of the plate is subjected to a distributed load $q = 96$ MPa. The displacement field is computed using the Navier-Cauchy equations, which can be simplified using the plane stress assumption as follows

\begin{equation}
\label{eq:navier_cauchy}
\begin{aligned}
    G(x)\left( \dfrac{\partial^2 u}{\partial x^2} + \dfrac{\partial^2 u}{\partial y^2} \right)
+ G(x) \dfrac{1 + \nu}{1 - \nu} \dfrac{\partial}{\partial x} \left( \dfrac{\partial u}{\partial x} + \dfrac{\partial v}{\partial y} \right)
 &= 0, \\
G(x)\left( \dfrac{\partial^2 v}{\partial x^2} + \dfrac{\partial^2 v}{\partial y^2} \right)
+ G(x) \dfrac{1 + \nu}{1 - \nu} \dfrac{\partial}{\partial y} \left( \dfrac{\partial u}{\partial x} + \dfrac{\partial v}{\partial y} \right)
 &= 0.
\end{aligned}
\end{equation}

where shear modulus \( G \) is defined as

\[
G(x) = \frac{E(x)}{2(1 + \nu)}
\]

Equation \eqref{eq:navier_cauchy} is solved for the displacement field $\mathbf{u}(x,y)$ using FEM. The plate is discretized into 2580 four-noded triangular elements. All FEM simulations are implemented on \texttt{FENICS} \cite{FENICS1,FENICS2}.

The limit state function is given as

\begin{equation}
    g(\mathbf{x}) = 320 - \max(\sigma_{11}(\mathbf{x}))
\end{equation}

where $\max(\sigma_{11}(\mathbf{x}))$ is the maximum principal stress in the plate for the given KL coefficients $\mathbf{x}$. The plate is assumed to fail if $\max(\sigma_{11}(\mathbf{x}))$ exceeds the yield strength of carbon steel (320 MPa). 

This example is a modification of a problem that has been studied in the past for reliability analysis, making it a suitable high-dimensional benchmark problem \citep{BUS_KLE, CE_plate}. We use this example to demonstrate the scalability of the DIRT framework for practical problems in structural reliability.

\begin{figure}
    \centering
    \includegraphics[width=\linewidth]{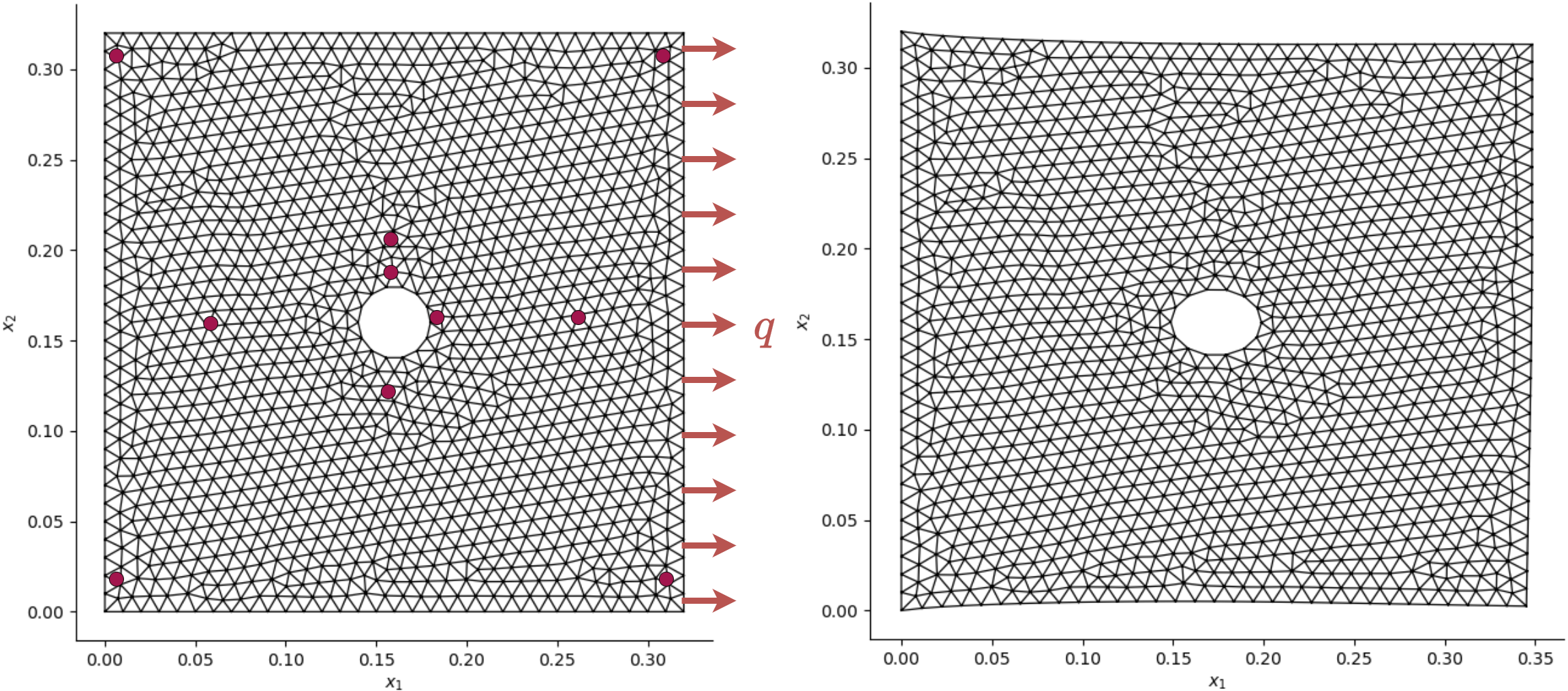}
    \caption{Plate with a circular hole subjected to traction loading.  Boundary conditions and sensor locations (left) and displaced mesh (right) }
    \label{fig:plateMesh}
\end{figure}

Axial strain data $\tilde{y} = [\epsilon_{x}, \epsilon_y]$ is observed at 20 locations on the plate as shown in Figure \ref{fig:plateMesh}, and a $2m$ dimensional Gaussian noise is added to these data points to introduce uncertainty in the system. The Gaussian noise is modeled using an exponential covariance function with mean zero, standard deviation $\sigma_\eta = 1\times10^{-4}$ and a correlation length $l_\eta = 0.08$ m. We also assume a correlation coefficient of 0.25 between the $x$ and $y$ directions of the measurement noises at the same location, and the measurement noise at different locations is independent. 

We represent the prior $E(x_1,x_2)$ as a lognormal random field with mean $\mu_E = 2\times10^5 $ MPa and standard deviation $\sigma_E = 3\times10^4$ MPa. The underlying Gaussian field is modeled using the exponential kernel with the autocovariance function given by \eqref{eq:kernel}
\begin{equation}
\label{eq:kernel}
    C(d) = \sigma^2_H \exp\left(-\frac{d}{l_c}\right).
\end{equation}

\begin{table}[htbp]
\centering
\begin{tabular}{cccccccc} 
\toprule
$\ell_c = 0.16$ & \multicolumn{2}{c}{DIRT} & \multicolumn{2}{c}{BUS-SuS} \\
\cmidrule(lr){2-3} \cmidrule(lr){4-5}
$M$ & $\Pfpostest$ & CoV (\# evals) & $\Pfpostest$ & CoV (\# evals) \\
\midrule
10  & $2.025 \times 10^{-7}$ & 0.018 (52676)   & $1.32 \times 10^{-7}$ & 0.190 (14000)\\
50  & $5.31 \times 10^{-6}$ & 0.025 (286676)  & $4.55 \times 10^{-6}$ & 0.133 (138700)\\
100 & $1.71 \times 10^{-5}$ & 0.017 (331968) & $3.00 \times 10^{-5}$ & 0.253 (272700)\\
250 & $5.57 \times 10^{-5}$ & 0.296 (1863368) & $6.37 \times 10^{-5}$ & 0.18 (675000)\\
\bottomrule
\end{tabular}
\vspace{20pt}
\caption{Comparison of DIRT and BUS-SuS estimates at $\ell_c = 0.16$, including coefficient of variation.}
\label{tab:dirt_bus_comparison_plate}
\end{table}


We conduct a parametric study to compare the performance of the DIRT framework against the BUS-SuS method. The study varies the number of KL coefficients $M$, which determines the effective dimensionality $D$ of the problem. The parameter settings are chosen as $l_c = 0.16 $ and $M \in \{10,\,50,\,100,\,250\}$. 

For each dimension, we estimate the posterior failure probability $\Pfpostest$ using both DIRT and BUS-SuS, and compute the CoV of these estimates based on 3 independent runs to assess the robustness of the DIRT estimator.  

For this example, tuning the optimal sigmoid-smoothing parameter $\gamma^\star$ was difficult since each estimation of $\Pfpostest$ required many forward evaluations of the FEM model. Therefore, $\gamma^\star$ is chosen by an empirical bias-vs-noise rule as follows. We evaluate $\Pfpostest^{(\gamma)}$ $N = 10$-times  on a grid $\Gamma = [50,100,200,500,1000]$ and take a sharp reference at $\gamma_{\max}=6000$ at $M=10$. 
Let $\mathrm{SE}(\gamma_{\max})$ denote the standard error at $\gamma_{\max}$. 
We select 
\begin{equation}
\label{eq:gamma_empirical}
\gamma^\star \;=\; \argmin_{\gamma\in\Gamma}
\Bigl|\; \bigl|\Pfpostest^{(\gamma)}-\Pfpostest^{(\gamma_{\max})})\bigr|
-\mathrm{SE}(\gamma_{\max}) \;\Bigr|
\end{equation}

where SE($\gamma_{\max}$) is the standard error of $\Pfpostest^{(\gamma_{\max})}$ defined as 

\begin{equation}
    SE(\gamma_{\max}) = \frac{\mathrm{std}(\Pfpostest^{(\gamma_{\max} )})}{ \sqrt{N}} \, .
\end{equation}

For this example, we used Eq. \eqref{eq:gamma_empirical} and set the value of $\gamma^\star = 200$ for all values of $M$. 

It is observed in Table~\ref{tab:dirt_bus_comparison_plate} that DIRT achieves estimates of $\Pfpostest$ that are comparable to BUS-SuS. For $M=10,50,$ and $100$, DIRT is able to achieve better values of CoV (10 times smaller than BUS-SuS). However, for $M = 250$, we can note that DIRT does not achieve the same reduction in CoV.  

\subsubsection{Estimation of True Young's Modulus Field}
\begin{figure}[ht]
  \centering
  \setlength{\tabcolsep}{3pt}
  \renewcommand{\arraystretch}{0.9}

  \begin{tabular}{@{}r *{3}{>{\centering\arraybackslash}m{0.29\textwidth}}@{}}
       & $M = 50$ & $M = 100$ & $M = 250$ \\[2pt]

    \makebox[0pt][r]{True} &
      \begin{subfigure}[t]{\linewidth}\includegraphics[width=\linewidth]{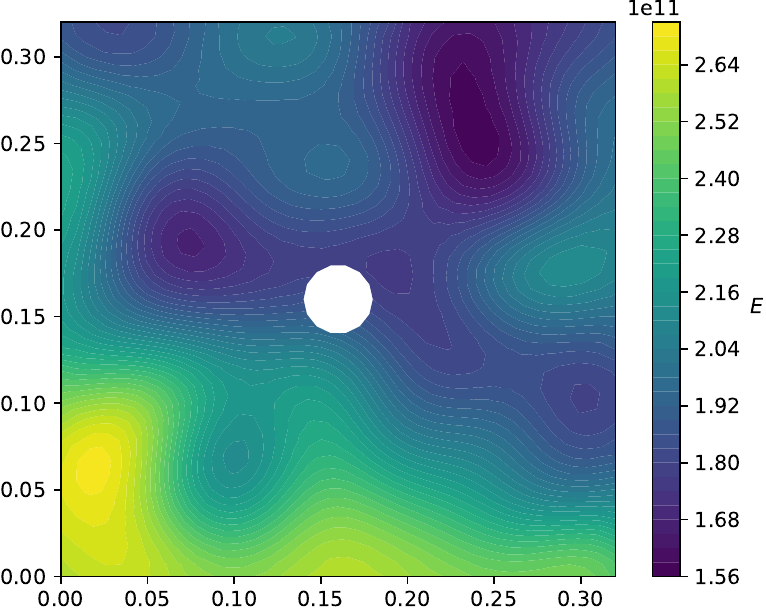}\end{subfigure} &
      \begin{subfigure}[t]{\linewidth}\includegraphics[width=\linewidth]{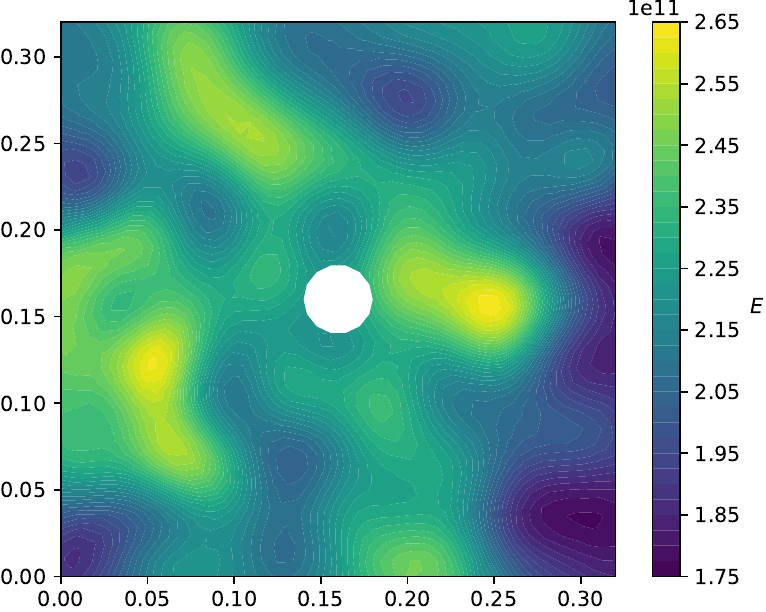}\end{subfigure} &
      \begin{subfigure}[t]{\linewidth}\includegraphics[width=\linewidth]{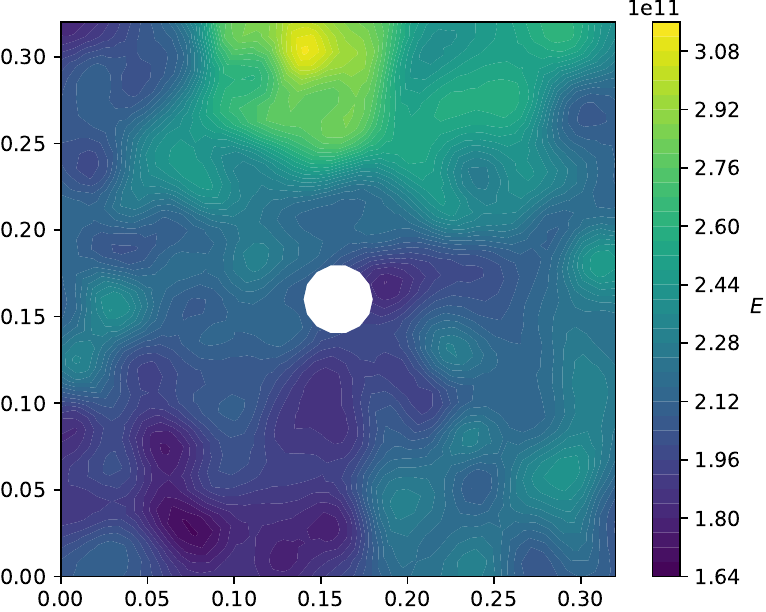}\end{subfigure} \\[6pt]

    \makebox[0pt][r]{Recovered} &
      \begin{subfigure}[t]{\linewidth}\includegraphics[width=\linewidth]{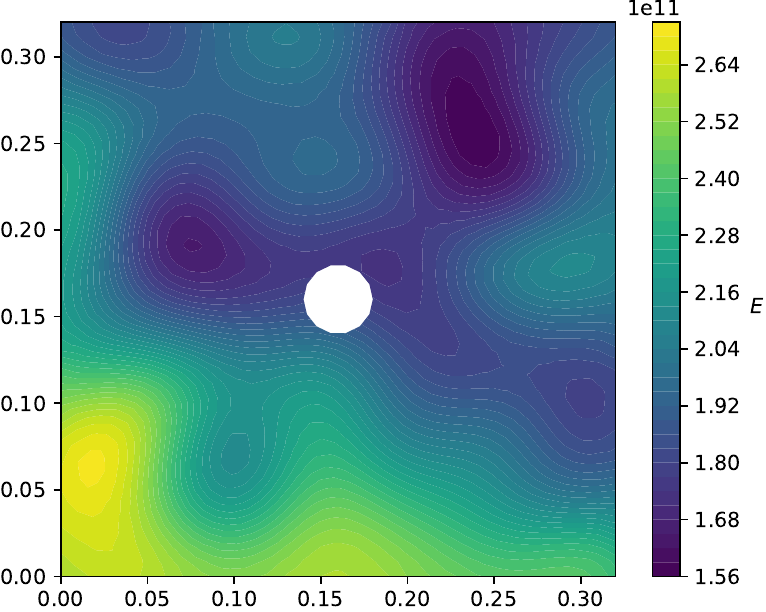}\end{subfigure} &
      \begin{subfigure}[t]{\linewidth}\includegraphics[width=\linewidth]{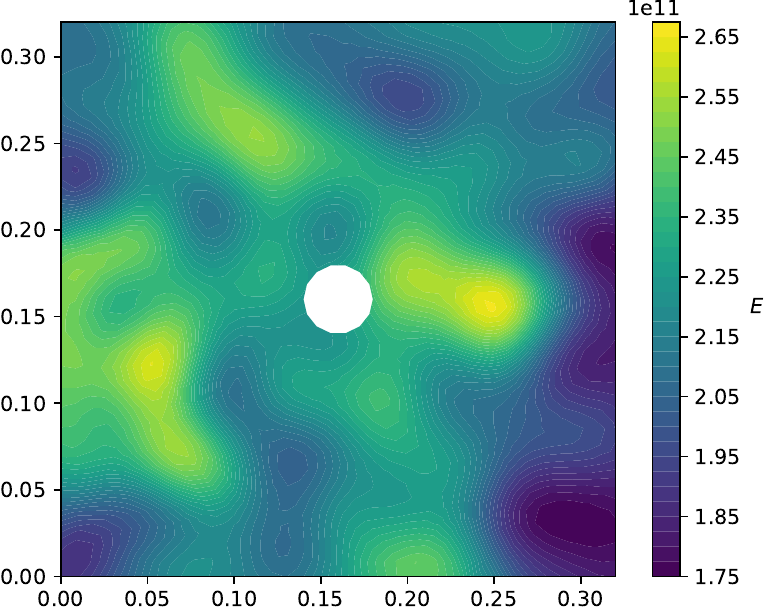}\end{subfigure} &
      \begin{subfigure}[t]{\linewidth}\includegraphics[width=\linewidth]{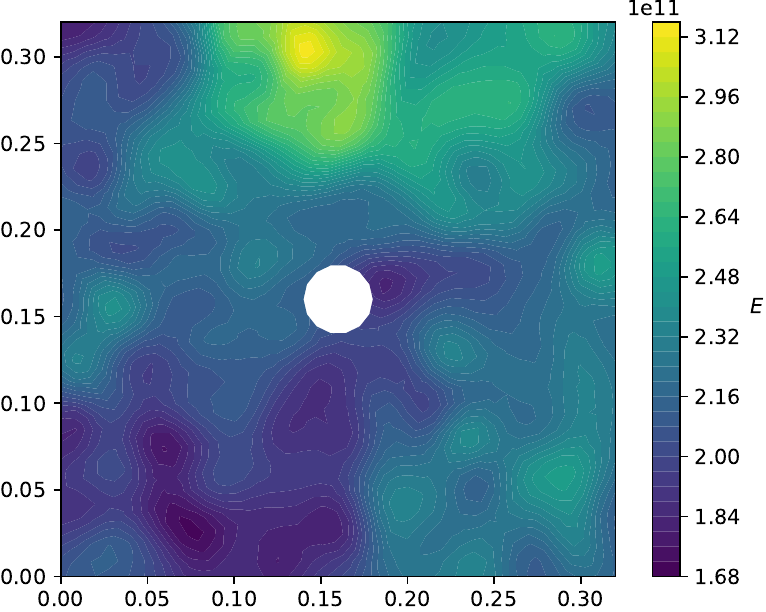}\end{subfigure} \\
  \end{tabular}

  \caption{Posterior Young's Modulus fields recovered from DIRT. Top row: ground truth fields; bottom row: posterior mean estimates obtained using DIRT with increasing parameter dimensionality.}
  \label{fig:E_post}
\end{figure}
Figure~\ref{fig:E_post} compares the true and recovered spatial distributions of the Young’s modulus field $E(x)$ for three values of the KL expansion dimension: $M = 50$, $100$, and $250$. The true fields are generated from a lognormal Gaussian process using the exponential kernel, and the posterior estimates are obtained using the DIRT framework.

Across all parameters, the posterior field recovered by DIRT exhibits high resemblance to the true field, capturing both the global structure and local variations of the modulus distribution. Notably, the posterior Young's modulus field approximation remains accurate even as the parameter dimension increases, despite the fact that the maximum TT rank used in the transport map approximation is fixed at $r = 4$. This demonstrates that for recovering the posterior Young's modulus field, DIRT generalizes well with respect to dimension and does not suffer from significant accuracy degradation with increasing $d$.

The recovered fields show consistent spatial agreement with the true fields, and the maximum relative error in any case remains within 10\%. The results validate the ability of DIRT to recover high-dimensional fields without requiring any significant increase in the rank of the TT representation.

\section{Summary and Conclusion}

In this work, we studied and evaluated a Tensor Train Decomposition-based Importance Sampling framework, known as DIRT \cite{DIRTRareEvent2022}, for estimating the probability of failure in structural reliability problems. The DIRT framework employs the TT-Cross algorithm to construct a low-rank approximation of the target probability density function, enabling efficient sampling even in high-dimensional settings.

Through a series of numerical experiments spanning low to high dimensions, we demonstrated that DIRT can outperform standard methods such as Bayesian Updating with the Structural Reliability method and the Cross-Entropy method in terms of estimator quality. Specifically, in some examples, DIRT is able to achieve significantly lower coefficients of variation of the failure probability estimates compared to these methods, indicating improved stability and reliability of the results. One notable advantage of the DIRT framework is its inherent ability to be parallelized. Since DIRT constructs a global surrogate of the posterior distribution via a deterministic low-rank approximation of the transport map, many forward model evaluations used to train this map can be generated independently. This contrasts with Bayesian Updating with Subset Simulation, which relies on conditional MCMC sampling within each subset level, making the algorithm inherently sequential and difficult to scale efficiently on parallel computing architectures. From a computing standpoint, this distinction yields some practical benefits since both transport maps can be approximated in parallel. Moreover, the layers of the TT approximation of the target PDF in DIRT are also independent and can be computed in parallel. This makes DIRT particularly suitable for computationally expensive forward models, such as linear and nonlinear finite element simulations. We also analyzed the computational cost of DIRT. The number of limit state function evaluations required by DIRT was shown to remain effectively independent of the rarity of the event, making it particularly suitable for estimating very small failure probabilities (e.g., \( \Pfpostest \approx 10^{-6} \)). However, the cost of DIRT scales approximately linearly with the problem dimensionality \( d \).

Despite these advantages, some limitations remain. For complex high-dimensional posterior distributions, a higher TT rank might be required to achieve sufficient accuracy, as shown in Section~3.4. This increases the number of forward model calls significantly, which can become prohibitive when the forward function is based on large-scale finite element models. Moreover, tuning the hyperparameters of DIRT (e.g., the TT rank \(r\), the smoothing parameter \(\gamma\), and the number of DIRT layers \(L\)) can be challenging in practice.

In summary, DIRT provides a scalable and parallel framework for high-dimensional reliability analysis and is particularly effective for problems involving extremely rare events. However, its efficiency is reduced for highly complex posterior distributions, where the need for large TT ranks might limit its applicability in computationally expensive structural reliability settings.

A promising direction for future research is the development of a multi-fidelity importance sampling \citep{MF_IS} framework for structural reliability analysis that builds upon the DIRT framework. As observed in Section 3.4, forward model evaluations (e.g., finite element simulations) are the major computational bottleneck in structural reliability.  A multi-fidelity approach has the potential to reduce the effective TT rank and substantially lower the number of expensive high-fidelity model evaluations required for training. In addition, adaptive strategies could be employed to allocate computational resources efficiently between fidelities \citep{MF1,MF2}. For instance, high-fidelity simulations may be focused near the failure domain, where accuracy is most critical for estimating small probabilities of failure, while low-fidelity models can be used to capture the global structure of the distribution.

\clearpage
\appendix

\section{Marginal and Cumulative Densities in TT Format}
\label{app1}
Here, we outline how the inverse Rosenblatt transformation is implemented in DIRT when the target density $\pi(\mathbf{x})$ is approximated in TT format. 
We adopt the conventional notation $f$ for probability density functions  
and $F$ for cumulative distribution functions.

\subsection*{A.1 Marginal and Conditional Densities}

Suppose the unnormalized density $\pi(\mathbf{x})$ is represented in TT form as
\begin{equation}
    \hat{\pi}(x_1,\ldots,x_d) = \pi^{(1)}(x_1)\,\pi^{(2)}(x_2)\cdots \pi^{(d)}(x_d), 
\end{equation}

where $\pi^{(k)}(x_k)$ are the TT cores defined in Section~2.1. 
The normalized density is then given by
\begin{equation}
\hat{f}_X(\mathbf{x}) = \frac{1}{\hat{c}}\,\hat{\pi}(x)\,\lambda(\mathbf{x}),
\quad \text{with} \quad 
\hat{c} = \int_{\mathcal{X}} \hat{\pi}(\mathbf{x})\,\lambda(\mathbf{x})\,d\mathbf{x}.
\end{equation}

The TT representation allows marginals to be computed by contracting the TT cores that 
do not correspond to the variables of interest. For example, the marginal density of the first $k$ variables is
\begin{equation}
\hat{f}_{X_{\leq k}}(x_{\leq k}) 
= \frac{1}{\hat{c}}\, \hat{\pi}_{\leq k}(x_{\leq k})
\prod_{i=1}^k \lambda_i(x_i),
\end{equation}
where $\hat{\pi}_{\leq k}$ denotes the partially contracted TT tensor 
with respect to the remaining $d-k$ dimensions. 

From this, the conditional density required for the IRT is
\begin{equation}
\hat{f}_{X_k|X_{<k}}(x_k|x_{<k}) 
= \frac{\hat{\pi}_{\leq k}(x_{\leq k})}{\hat{\pi}_{\leq k-1}(x_{<k})}\,\lambda_k(x_k).
\end{equation}

The corresponding conditional CDF is
\begin{equation}
\hat{F}_{X_k|X_{<k}}(x_k|x_{<k}) 
= \int_{-\infty}^{x_k} \hat{f}_{X_k|X_{<k}}(y_k|x_{<k})\,dy_k.
\end{equation}

Given an independent uniform $\mathbf{u}=(u_1,\ldots,u_d)\sim U([0,1]^d)$,  samples from the target distribution are obtained by sequential inversion:
\begin{equation}
x_1 = \hat{F}_{X_1}^{-1}(u_1), \quad
x_2 = \hat{F}_{X_2|X_1}^{-1}(u_2|x_1), \quad \ldots, \quad
x_d = \hat{F}_{X_d|X_{<d}}^{-1}(u_d|x_{<d}).
\end{equation}

\clearpage

\bibliography{references}

\end{document}